\documentclass[%
 aip,
 amsmath,amssymb,
 reprint,%
]{revtex4-1}

\usepackage{graphicx}
\usepackage{dcolumn}
\usepackage{bm}

\usepackage[utf8]{inputenc}
\usepackage[T1]{fontenc}
\usepackage{mathptmx}
\usepackage{etoolbox}
\usepackage{xcolor}

\makeatletter
\def\@email#1#2{%
 \endgroup
 \patchcmd{\titleblock@produce}
  {\frontmatter@RRAPformat}
  {\frontmatter@RRAPformat{\produce@RRAP{*#1\href{mailto:#2}{#2}}}\frontmatter@RRAPformat}
  {}{}
}%
\makeatother
\begin{document}

\preprint{AIP/123-QED}

\title[]{Controlling the interaction of tightly focused 10-PW class lasers with multicomponent plasma via target parameters: optimization of electron-positron pair and $\gamma$-photon sources}
\author{A.~V.~Bashinov}%
 \email{bashinov@ipfran.ru}
 \affiliation{%
A.V. Gaponov-Grekhov Institute of Applied Physics of the Russian Academy of Sciences, Nizhny Novgorod 603950, Russia
}%

\author{E.~S.~Efimenko}
 \affiliation{%
 A.V. Gaponov-Grekhov Institute of Applied Physics of the Russian Academy of Sciences, Nizhny Novgorod 603950, Russia}

\author{A.~A.~Muraviev}
\affiliation{%
A.V. Gaponov-Grekhov Institute of Applied Physics of the Russian Academy of Sciences, Nizhny Novgorod 603950, Russia
}%

\author{E.~A.~Panova}
\affiliation{%
Lobachevsky State University of Nizhni Novgorod, Nizhny Novgorod 603950, Russia
}%

\author{V.~D.~Volokitin}
\affiliation{%
Lobachevsky State University of Nizhni Novgorod, Nizhny Novgorod 603950, Russia
}%

\author{I.~B.~Meyerov}
\affiliation{%
Lobachevsky State University of Nizhni Novgorod, Nizhny Novgorod 603950, Russia
}%

\author{A.~V.~Kim}
\affiliation{%
A.V. Gaponov-Grekhov Institute of Applied Physics of the Russian Academy of Sciences, Nizhny Novgorod 603950, Russia
}%

\date{\today}

\begin{abstract}
The interaction of multipetawatt lasers with plasma is a complex multiparameter problem, providing a wide field for fundamental research and opening up great opportunities for creating unique sources of high-energy electrons and positrons, dense pair plasma, and $\gamma$-photons. However, to achieve high efficiency of such a source, it is necessary to use targets with optimized parameters, primarily density and size, for the given laser parameters. With the use of 3D QED-PIC modeling it is shown that, when targets whose size is comparable with the laser wavelength are irradiated by laser beams with a total power of several tens of PW, the total initial number of target electrons may be regarded to be the similarity parameter of laser-plasma interaction. In practice, this can significantly simplify the selection of the targets needed for controlling the interaction and, accordingly, for achieving the specified parameters of the developed electron-positron plasma and $\gamma$-photon sources. Based on the similarity parameter, various laser-plasma interaction modes are identified, the necessary conditions for their launch are determined, and the properties of the pair particle and $\gamma$-photon source are revealed. Moreover, qualitative estimates of the quantitative and energy characteristics of such a source are obtained, allowing it to be optimized for various laser beam configurations.
\end{abstract}

\maketitle

\section{\label{sec:Intro}Introduction}
Lasers of 10-PW power are at the verge of operation and laser systems of 100-PW or exawatt level are well under way \cite{Danson_HPLSE2019,Khazanov_HPLSE2023}. Tight focusing of such powerful laser beams can enable  new regimes of laser plasma interaction. First, radiative effects come into play in extreme fields \cite{Gonoskov_RMP2022,Popruzhenko_UFN2023,Blackburn_RMPP2020}. It was experimentally proved that the motion of accelerated electrons in the fields of petawatt lasers may be notably affected by photon recoil \cite{Cole_PRX2018,Poder_PRX2018}. A number of theoretical investigations, in turn, showed that electrons at intensities $\gtrsim5\times10^{23}\mathrm{~W~cm^{-2}} $ can generate gamma-photons with high efficiency \cite{Koga_PP2005,Bashinov_PP2013,Nerush_PP2014,Zhang_NJP2015,Fedotov_PRA2014,Nakamura_PRL2012,Brady_PRL2012,Ji_PRL2014,Gonoskov_PRL2014,Chang_SR2017}, thus contributing to absorption in collisionless plasma. This may qualitatively modify plasma properties \cite{Liseykina_NJP2016,Bashinov_PP2013,Gelfer_SR2018,Gelfer_PRE2020}.

Second, abundant decay of gamma-photons into electron-positron pairs at intensities $>10^{24}\mathrm{~W~cm^{-2}}$ can feed a quantum electrodynamic cascade (QED) \cite{Bell_PRL2008,Nerush_PRL2011,Fedotov_PRL2010,Bulanov_PRL2010S,Grismayer_PRE2017,Mironov_PRA2021}. It was shown that tight focusing of laser radiation of about 10-PW level leads to the generation of overcritical e$^-$-e$+$ plasma \cite{Gelfer_PRA2015,Gonoskov_PRX2017,Jirka_SR2017}. As a result, the QED cascade can change the plasma composition drastically and, consequently, the character of plasma dynamics and plasma response \cite{Edwards_PRL2016,Tsytovich_CPPCF1978,Helander_PRL2014,Qu_PPCF2023,Efimenko_SR2018,Efimenko_PRE2019,Muraviev_JETPL2015,Samsonov_SR2019,Grismayer_PP2016}. 

However, in order to trigger this or that regime it is necessary to choose suitable target parameters such as its composition, density, size and shape. Recent studies partially tackled this problem. It was found how initial plasma density can facilitate or suppress $\gamma$-photon generation or QED cascade development \cite{Nakamura_PRL2012,Nerush_PP2014,Bashinov_PP2013,Jirka_SR2017,Slade_NJP2019}. Also it was shown that in the case of a  gas target it is possible to govern the QED cascade development using target composition \cite{Artemenko_PRA2017,Tamburini_SR2017}.

Thus for upcoming experiments it is highly important to investigate the mutiparametric problem of the interaction of multipetawatt lasers with matter depending on laser power, intensity, duration, configuration of focusing, as well as on target composition, size, shape, and density. In addition to fundamental knowledge, this may provide a basis for the creation of controllable sources of $\gamma$-photons, dense pair plasma, highly energetic beams of e$^-$ and e$^+$, and so on.

Since laser radiation parameters are mainly determined by planned facilities, our paper is devoted to the influence of target parameters on the interaction regimes. We have chosen not so widely studied but one of basic field configurations, that is a dipole field configuration \cite{Gonoskov_PRL2014,Gonoskov_PRA2012} underlying the XCELS project \cite{Khazanov_HPLSE2023}. This field configuration maximizes field strength at a given total laser power, thus facilitating photon emission and pair production \cite{Gonoskov_PRX2017}, and allows obtaining unique extreme plasma dynamics \cite{Efimenko_SR2018,Efimenko_PRE2019}. We consider both an ideal incoming dipole wave and a relevant to future experiments multibeam configuration mimicking a dipole wave with high accuracy.

We emphasize that it is not so much density as a combination of target parameters that determines the interaction regimes, especially in case of a target irradiated by counterpropagating laser pulses. Despite our study concerns mainly spheres of various densities and sizes as targets, the obtained results are generalized to other shapes as well. Also, we specify requirements for target density and size in  order to initiate a QED cascade and to control the properties of $\gamma$-photons, electron and positron beams generated as a result of laser irradiation of the target.

The structure of this paper is as follows. Since the main instrument of our investigation is PIC-simulation, to begin with we present our numerical setup in Sec.~\ref{sec:NumSetup}. Then in Sec.~\ref{sec:ideal} we analyze the interaction of an ideal e-dipole wave with targets. We describe the interaction regimes and emerging basic plasma-field structures, determine target parameters corresponding to different regimes and consider the properties of accelerated electrons and positrons escaping the focal region and the properties of the generated $\gamma$-photons. Based on numerical simulations we obtain qualitative estimates of the energy characteristics of the generated $\gamma$-photons and accelerated particles. In Sec.~\ref{sec:multi} we analyze how robust the interaction considered in the previous section is. For this we consider the interaction of targets with a number of tightly focused laser beams configured as a dipole wave and show how the properties of accelerated positrons and generated $\gamma$-photons are modified with a change in laser power, pulse duration, number of laser beams, and their polarization. Finally, in Sec.~\ref{sec:con} we summarize the obtained results.

\section{\label{sec:NumSetup}Numerical setup}
The extreme light-matter interaction is strongly nonlinear \cite{Mourou_RMP2006} and its study demands considering  various complex QED processes in strong fields \cite{Piazza_RMP2012}, primarily, $\gamma$-photon emission and photon decay into pairs. One of the key instruments of such investigations is QED-PIC simulation\cite{Ridgers_JCP2014,Grismayer_PP2016,Derouillat_CPC2018,Fedeli_NJP2022,Montefiori_CPC2023}. We use the advanced highly-optimized QED-PIC code PICADOR \cite{Bastrakov_JCS2012} based on efficient algorithms to model QED processes \cite{Gonoskov_PRE2015,Muraviev_CPC2021,Volokitin_JCS2023,Panova_2021}. The emission and decay of $\gamma$-photons are determined by spectral probability densities obtained within the framework of quantum electrodynamics in the Local Constant Field Approximation (LCFA) \cite{Ritus_85,Baier_NPB1989}.

Following the multipetawatt laser projects, especially XCELS \cite{Khazanov_HPLSE2023}, the wavelength $\lambda$ is chosen to be $0.9~\mu$m, the corresponding wave period $T$ is approximately 3~fs and the frequency is $\omega=2.1\times10^{15}$. The temporal shape of the pulse is Gaussian with pulse duration $\tau=5T=15$~fs (FWHM for intensity). This duration corresponds to the minimum achievable one at the prospective XCELS facility. We choose the total laser power P=30 PW. Laser pulses of such power can trigger vacuum breakdown and the formation of both current sheets and pinch structures \cite{Efimenko_SR2018,Efimenko_PRE2019}.

We consider the dipole field configuration that can be mimicked by a number of laser beams, which allows maximizing the field strength at a given laser power \cite{Gonoskov_PRA2012,Gonoskov_PRL2014}. We examine and compare the numerical results obtained in both cases: an ideal dipole wave and multibeam setups. An ideal dipole wave is generated by means of the spatio-temporal distribution of currents in the boundary planes forming a spherical incoming phase front. The applicability of this approach was verified in our earlier works  \cite{Gonoskov_PRX2017,Efimenko_PRE2019,Efimenko_SR2018}. The central point, where the field has maximum amplitude, coincides with the center of the computational domain and has coordinates (0,0,0). The  field symmetry axis is the $z$ axis. The strongest field in focus is directed along this axis (electric field in case of e-dipole wave or magnetic field in case of m-dipole wave). In order to simulate a multibeam setup a special program module within PICADOR code was developed \cite{Panova_2021}. This module allows setting all the properties of laser beams: power $P$, spatial and temporal envelope, laser wavelength $\lambda$ and frequency $\omega$, $\tau$, angle of propagation with respect to the coordinate system, angle of focusing, and position of the focus. In the case of the multibeam setup, all laser pulses are focused on the coordinate origin and arrive to this point synchronously. Here we consider two multibeam setups: 12 beams arranged in two belts and 6 beams arranged in one belt as proposed in Ref.~\cite{Gonoskov_PRL2014}. In addition, a series of simulations for a duration of 30 fs and a power of 15~PW have been performed to evaluate the effect of pulse duration and laser power on laser-plasma dynamics.

In all performed simulations the computational domain is a cube with a side of 8$\lambda$ and includes 512 cells in each directions $x$, $y$ and $z$. So, the spatial resolution is $\lambda/64$, while the temporal resolution is $T/150$. For resolving QED processes, the step is automatically subdivided within the developed Optimized Event Generator \cite{Volokitin_JCS2023}. The total simulation time ranges from $40T$ to $53T$ depending on the pulse duration.

Targets have a spherical shape. Their radius $R_0$ ranges form $1\mu m$ to $3\mu m$. The target density $n_0$ varies logarithmically from $0.001n_{cr}$ to $1000n_{cr}$, where $n_{cr}=m\omega^2/(4\pi e^2)\approx1.4\times10^{21}~\mathrm{cm}^{-3}$ is the critical plasma density, $m$ is the electron mass, $e$ is the elementary charge. This density range (from 10$^{18}$ cm$^{-3}$ to 10$^{24}$ cm$^{-3}$ in dimensional units) covers the entire range of available target materials from aerogels to high-Z materials. Target ionization is not considered in this work. Due to the extremely high intensity of laser radiation and small target sizes we assume that the target is completely ionized by the leading edge of the laser pulse. So, we  consider ions with charge $e$ and mass $2m_p$, where $m_p$ is the proton mass. Using more realistic heavy ions in simulations does not affect the results qualitatively but may quantitatively influence some aspects of dynamics when ion mobility is important, especially for high-Z materials. The initial number of macroparticles (electrons and ions) is $2\times10^{7}$ of each particle type.

In total we have performed around 10 series of simulations depending on target size, field structure, and each series included 9 simulations with different $n_0$. So, around 100 simulations have been carried out. 10 to 14 hours of computing  were needed to complete the simulation. The analysis of the results is based on the distributions of fields and particles (electrons, positrons) in the focal region, on the spectra of particles and $\gamma$-photons, as well as on the efficiency of conversion of laser energy to photon energy and to energy of accelerated electrons and positrons. 

\section{\label{sec:ideal}Ideal dipole wave}

We will first consider an ideal e-dipole wave, in which  vacuum breakdown was previously studied in \cite{Gonoskov_PRX2017,Efimenko_SR2018,Efimenko_PRE2019}. Here we focus on all basic regimes of the interaction of an e-dipole wave with different targets. For all considered target parameters, initially incident laser radiation compresses a target, but how strong this compression occurs, whether strong laser fields penetrate into plasma, whether vacuum breakdown can be triggered, and which basic plasma-field structure can be formed highly depend not only on the initial density but also and more importantly on the combination of the initial density and size of the target.

In our simulations we distinguish three basic plasma-field structures shown in Fig.~\ref{fig:structs}, that will be described in section \ref{subsec:plasmastruct}. These basic structures can be formed at all considered target radii; however, the necessary threshold of the initial density is the lower, the larger the target radius. This fact confirms that some combinations of density $n_0$ and radius $R_0$ of the target determine the character of laser-plasma interaction.  Firstly, in the next section we will differentiate between these structures by the initial total quantity of electrons within target $N_0$ which is the simplest dimensionless combination of $n_0$ and $R_0$. Later in the text, the use of $N_0$ will be justified. More rigorously, the relationship of threshold target densities and their radius will be analyzed in section \ref{subsec:dependencetarget}. A simplified translation of $N_0$ to $n_0$ and back is
\begin{eqnarray}
    &n_0=1.7\times10^{-10}N_0/R^3_{\mu m},\label{Eq:n0N0}\\
    &N_0=6n_0R^3_{\mu m}10^{9},\label{Eq:N0n0},
\end{eqnarray}
where $R_{\mu m}$ is the initial radius of the target in microns.

Note that, in the following figures and below in the text, the densities are normalized by the non-relativistic critical density $n_{cr}$ and the fields are normalized to the relativistic field $E_{rel}=m\omega c/e\approx1.2\times10^8$~statV~cm$^{-1}$, where $c$ is the speed of light. With such normalization, maximal amplitudes of dimensionless electric $E_a$ and magnetic $B_a$ fields of the e-dipole wave in vacuum are equal to 
\begin{eqnarray}
    E_a\approx780\sqrt{P_{PW}}\approx4300,\label{Eq:E_P}\\
    B_a\approx510\sqrt{P_{PW}}\approx2800,\label{Eq:B_P}
\end{eqnarray}
where $P_{PW}=P/1\mathrm{PW}=30$. Due to the strong plasma reaction on the fields, in the basic structures the field amplitudes are less (or much less) than $E_a$ and $B_a$.

\subsection{\label{subsec:plasmastruct}Plasma-field structures}

\begin{figure*}
	\includegraphics[width=1\linewidth]{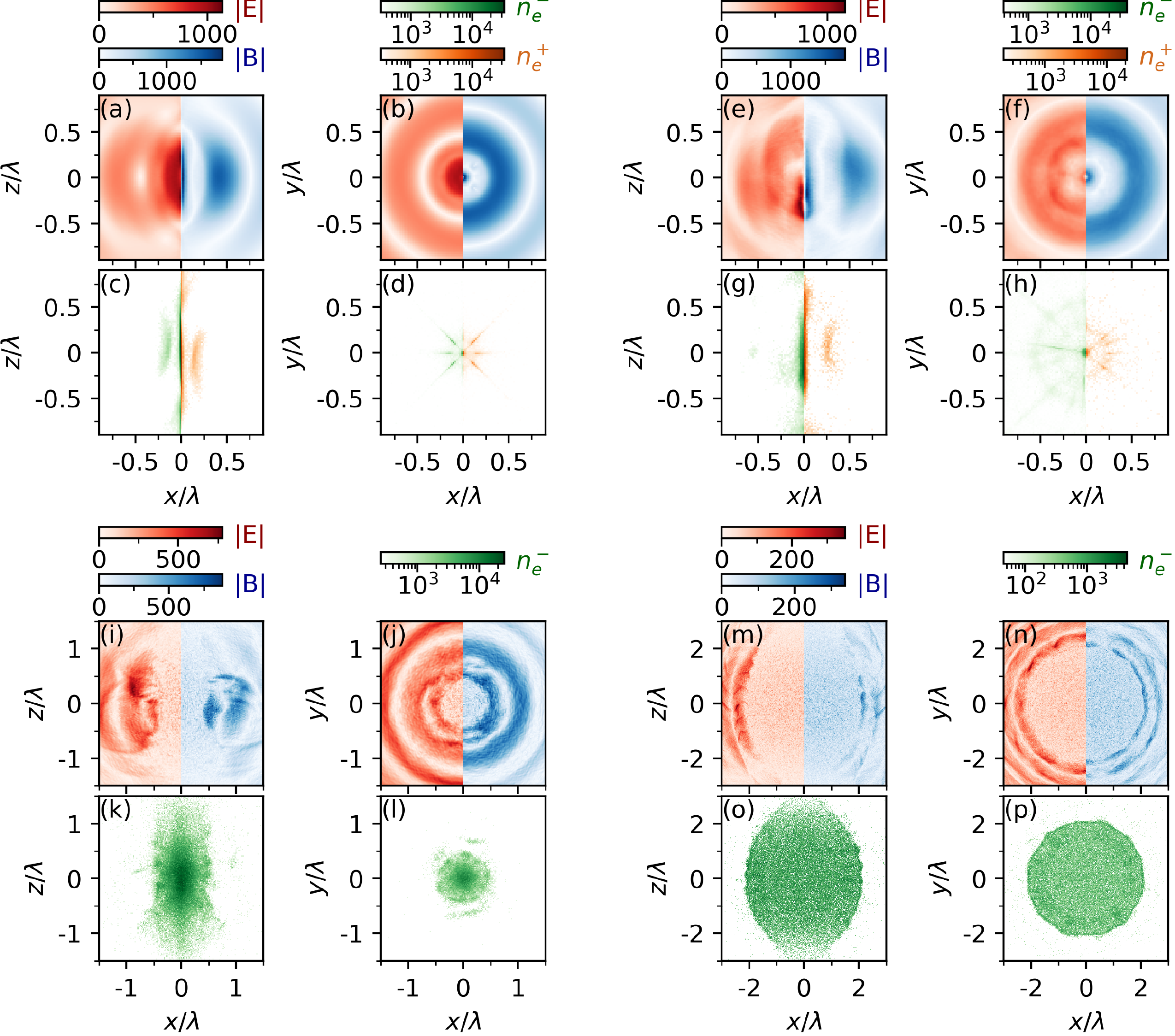}
	\caption{\label{fig:structs} Basic plasma-field structures formed as a result of irradiation of a spherical target by e-dipole wave. Examples of basic structures obtained at $R_0=3\mu$m and $n_0=0.001$ (a)-(d), at $R_0=2\mu$m and $n_0=30$ (e)-(h), at $R_0=1\mu$m and $n_0=1000$ (i)-(l), at $R_0=2\mu$m and $n_0=1000$ (m)-(p). The fields are shown in $xz$ plane in panels (a), (e), (i), (m) and in $xy$  plane in panels (b), (f), (j), (n). The left and right sides of these panels correspond to electric (in shades of red) and magnetic (in shades of blue) fields, respectively. Electron (left side of the panel, in shades of green) and positron (right side of the panel, in shades of orange) distributions are presented in $xz$ plane in panels (c), (g), (k), (o) and in $xy$ plane in panels (d), (h), (l), (p). In panels (k), (l), (o), (p) only electrons are depicted because the quantity of positrons in these structures is negligible and they do not form any structure.}
\end{figure*}

The first basic structure (see Figs.~\ref{fig:structs}(a)-(d)) can be observed at a "relatively small" initial total quantity $N_0$ of target electrons. The incident laser radiation significantly presses the initial target to the $z$ axis and the strong field penetrates through the entire plasma distribution. The initially ionized target is significantly destroyed by incoming laser radiation and the remaining focal electron-ion plasma ensures seeds for QED cascade development. Since $P>10$~PW,  the vacuum breaks down, and electron-positron pairs are generated abundantly \cite{Gonoskov_PRX2017}. In the field of the laser wave e$^{-}$-e$^{+}$ plasma self-compresses, resulting in the first basic structure. After fast exponential growth, the quantity of pairs saturates at the level of $10^{11}$ (the escape of pairs from the focus is balanced by their generation \cite{Efimenko_PRE2019}). This quantity is approximately kept until the laser fields weaken due to a finite pulse duration and further this structure decays. 

So, quasi-stationary plasma distribution is a very thin overcritical column and thin overcritical current sheets elongated along the $z$ axis (see Figs.~\ref{fig:structs}(c) and (d)) previously studied in Refs. \cite{Efimenko_SR2018,Efimenko_PRE2019}. The plasma column has a radius of $\approx0.016\lambda$ for the given resolution and a height of about $\lambda$. The current sheets are shifted by $0.15\lambda$ from the $z$ axis in radial direction (in the sense of cylindrical coordinate system), they have a height of $0.4\lambda$, radial thickness of about $0.07\lambda$ and are very thin in azimuthal direction with thickness around $0.016\lambda$. Note that so far it has been impossible to resolve these extremely compressed e$^-$-e$^+$ structures \cite{Efimenko_SR2018,Efimenko_PRE2019}. The field distributions  (Figs.~\ref{fig:structs}(a) and (b)) demonstrate the field decay within the overcritical plasma structures and reveal regions of strong enhancement of inductive fields in the vicinity of the structures. These regions are visually clearer  for the magnetic field because these structures are located around the magnetic field node of the standing wave (the $z$ axis). Since mainly e$^-$-e$^+$ plasma generated because of the vacuum breakdown determines the plasma-field dynamics and the formation of the first basic structure, the regime of target irradiation by the wave leading to the formation of this structure can be called {\it the regime of $e^-$-$e^+$ plasma self-compression}. 

Current sheet formation and plasma pinching are observed simultaneously, although the first regime should be changed by the second one at $P\approx20$~PW \cite{Efimenko_PRE2019}. Such a combination of regimes is mainly explained by a smooth laser envelope, while a half-infinite wave with a sharp front was considered in Ref.~\cite{Efimenko_PRE2019}. Also, since we consider a laser pulse of finite duration, the basic structure not always has sufficient time to become quasi-stationary. While its geometrical sizes are almost the same, with the growth of $N_0$ the fraction of particles involved in pinching instead of the current sheet formation increases. Besides,  the maximum density of this basic structure rises until it saturates at the level of $n_{e^-}^\mathrm{max}\approx n_{e^+}^\mathrm{max}\approx\times10^5$ (in dimensional units $\approx10^{26}$~cm$^{-3}$).

With a further growth in $N_0$ several transformations occur. The plasma column radius increases from $0.016\lambda$ to $0.02-0.03\lambda$, the current sheets become wider in radial direction and the second basic plasma-field structure appears. This structure is observed at "moderate" $N_0$ (Figs.~\ref{fig:structs}(e)-(h)). According to the analysis of the plasma-field distribution of each series of simulations, this transformation occurs within the density ranges $30<n_0<50$, $1<n_0<10$, $1<n_0<10$ (or within the ranges $1.8\times10^{11}<N_0<3\times10^{11}$, $4.8\times10^{10}<N_0<4.8\times10^{11}$, $1.6\times10^{11}<N_0<1.6\times10^{12}$) for targets with radii $R_0=1,~2,~3~\mu$m, respectively.

Although the second basic plasma-field structure is like the first one, it has several distinctions. Larger $N_0$ does not allow laser radiation to destroy the structure significantly and compress the target as much as at lower $N_0$. Nevertheless, the strong field can still penetrate through the entire plasma distribution, vacuum breaks down, and pinching as well as current sheet formation are observed. So, the mixture of e$^-$-e$^+$ and electron-ion plasmas determines the interaction with laser radiation. As a result, the plasma column due to pinching is thicker (see Fig.~\ref{fig:structs}(g)) and there emerge electron-ion current sheets (they are approximately twice as far from the $z$ axis as $e^-$-$e^+$ current sheets, compare the left and right sides in Fig.~\ref{fig:structs}(h)). A larger structure causes the decay of wave fields within a larger region, where inductive fields more clearly exceed the wave fields (see Figs.~\ref{fig:structs}(e) and (f)). One more distinction is that the electron density is noticeably higher than the positron density. This regime of target irradiation by laser radiation can be called {\it a regime of multicomponent plasma self-compression}. The dynamics of the quantity of electrons and positrons in focus is similar to that in {\it the regime of $e^-$-$e^+$ plasma self-compression}.

A further growth of $N_0$ leads to an increase in the fraction of electrons in current sheets, a decrease in the plasma column density and its unstable behavior. This column can randomly shift as a whole from the $z$ axis at a distance of $~0.1\lambda$ (around the displacement of current sheets), bend and join the current sheets. Such behavior is the evidence of the transformation of the second basic plasma-field structure to the third one.

The third basic structure is formed at "high" $N_0$. Simulations show that the transformation of the second basic structure to the third one for targets with $R_0=1,~2,~3~\mu$m should be within the intervals $100<n_0<1000$ (or according to (\ref{Eq:n0N0}) at $6\times10^{11}<N_0<6\times10^{12}$), $30<n_0<50$ ($1.4\times10^{12}<N_0<2.4\times10^{12}$), and $10<n_0<30$ ($1.6\times10^{12}<N_0<4.9\times10^{12}$), respectively. There is no unified single structure. Different  variations are possible (see, for example, Figs.~\ref{fig:structs}(i)-(l) or Figs.~\ref{fig:structs}(m)-(p)), but their common feature is that laser radiation only partially compresses the target and a strong field cannot penetrate deep into plasma. The spatial scale of the plasma distribution is comparable with the wavelength and is larger or much larger than the skin depth (Figs.~\ref{fig:structs}(k), (l) and (o), (p)). Due to weak target compression and field configuration of the e-dipole wave corresponding to utmost focusing, the strongest fields of the third structure are weaker or much weaker than the ones in the two previous plasma-field structures (compare Figs.~\ref{fig:structs}(i), (j), (m), (n) with Figs.~\ref{fig:structs}(a), (b), (e), (f)). Therefore, pair generation vanishes, and we do not depict positron distribution in Figs.~\ref{fig:structs}(k), (l), (o), (p). The character of laser-plasma interaction is determined only by the electron-ion plasma. This regime of laser-target interaction can be called {\it weak laser compression of a target}.

Note also that the interaction of laser radiation with solid density targets is accompanied by an instability development, which results in plasma density modulation within the skin layer in both azimuthal and polar directions (see Figs.~\ref{fig:structs}(k), (l), (o) and (p)). However, analysis of this instability is beyond the scope of this paper.

The absence of a unified basic plasma-field structure in the third case is explained by weak laser compression of a dense enough target. Depending on the target radius and density, the compression length may be much less than the target radius. So, the evolution of compression is generally unique at "high" $N_0$. Contrariwise, when the first and the second basic structures are being formed, plasma is strongly compressed into focus (to the $z$ axis), where its distribution acquires a cylindrical shape. Such a shape is caused by the field structure. Thus, the plasma-field distribution evolves through comparable structures or distributions in focus. Therefore, the geometrical dimensions of these structures are more or less identical at different target radii and densities, while the  plasma density can vary due to the finite duration of the laser pulse and insufficient time for the structure to become quasi-stationary. However, if an intermediate plasma-field distribution is similar in all respects, even for a target with different parameters, then the formed basic structure will also be quantitatively similar.

When the first or the second basic structure is being formed, not all electrons of target are dragged  into focus by laser radiation: some particles can immediately leave the strong field region due to the strong field inhomogeneity, while another part of particles can be scattered by diverging laser radiation (passed through the focus), lose the gained energy due to radiative effects \cite{Fedotov_PRA2014}, and fail to reach the region of the strongest fields. The quantity of electrons gathered in focus $N_c$ usually depends on the envelope, wave power \cite{Bashinov_QE2013}, target parameters and demands rigorous investigations. Nevertheless, the simulations for the considered target and wave parameters show (confirmatory results of the simulation are available in Sec.~\ref{sec:ideal}) that this quantity in case of a target with size about $\lambda$ is primarily proportional to $N_0$ and the proportionality constant is $f_{darg}\approx0.1$. The main consequence of this is that, in the zeroth approximation for the first and the second structures, the similarity parameter is $N_0$. In other words, if the targets have similar $N_0$, then not only the geometry of plasma-field structure but also the plasma density and field strengths are similar. Further analysis is mainly based on the parameter $N_0$.

\subsection{\label{subsec:dependencetarget} Dependence of basic structures on $N_0$}

The description of basic structures in the previous section reveals that the initial compression of a target by incident laser radiation determines which structure may be formed. If the initial target is compressed and the fields penetrate into plasma, then the first or the second basic structure is realized, when abundant e$^-$-e$^+$ pair production occurs, and there arise plasma pinching and current sheet formation. Otherwise, depending on the initial target density and size, a different variation of the third basic structure is possible. Their common feature is that the spatial size of the structure is comparable to the laser wavelength and (much) larger than the skin depth.

Thus, the threshold plasma-field distribution formed due to target compression and distinguishing the second and the third basic structures is similar to the plasma cylinder with the skin depth $\delta$ equal to its radius $r_{cth}$:
\begin{equation}
    \label{Eq:targetcondition}
    r_{cth}=\delta=1/k_\mathrm{Im},
\end{equation}
where $k_\mathrm{Im}$ is the imaginary part of the wave number. The height of this cylinder is the diameter of the focal spot (measured over the field amplitude) $h_{c}\approx2w_0\approx0.9\lambda$, so the volume of the plasma cylinder is $V_{cth}=\pi r_{cth}^2 h_{c}$.

The fields can still penetrate throughout the plasma cylinder, but the vacuum breakdown does not change the quantity of electrons, since their generation is balanced by their escape. As discussed above, the threshold quantity of electrons gathered in focus in the considered cylinder is $N_{cth}=f_{drag}N_{0th}$, where $N_{0th}$ is the threshold initial quantity of target electrons. Then the characteristic average density of this plasma cylinder is:
\begin{equation}
    \label{Eq:nc}
    n_{cth}=\frac{N_{cth}}{n_{cr}V_{cth}}\approx \frac{f_{drag}N_{0th}}{n_{cr}V_{cth}}\approx \frac{0.13n_{0th}R_0^3}{n_{cr}r_{cth}^2h_{c}},
\end{equation}
where $n_{0th}$ is the threshold initial target density.

The strong electric field of the wave accelerates the electrons within this cylinder to ultrarelativistic energies. Although ultrarelativistic plasma is mainly collisionless, the dielectric constant $\varepsilon$ of plasma in the  ultrarelativistic laser field may be (in dimensionless fields $\gtrsim500$) like in the case of collisions due to photon emission by electrons \cite{Bashinov_PP2013}:
\begin{equation}
    \label{Eq:epsilon}
    \varepsilon=1-\frac{n_{cth}}{\gamma\left(1-i\nu/\omega\right)},
\end{equation}
where $\gamma$ is the Lorentz-factor of the electrons, $\nu$ is the analog of effective collisional frequency due to strong perturbation of electron motion by $\gamma$-photon emission. Since at $P=30$~PW the particles move in the anomalous radiative trapping (ART) regime \cite{Gonoskov_PRL2014}, strong photon recoil during particle motion occurs twice per wave period, so 
\begin{equation}
    \label{Eq:nuomega}
    \nu/\omega\approx2.
\end{equation}

As shown by the previous studies in the e-dipole wave \cite{Efimenko_SR2018,Efimenko_PRE2019},  in a  wide range of powers higher than 10~PW, the  electron energy slightly depends on power and the average Lorentz-factor is roughly 
\begin{equation}
    \label{Eq:estgamma}
    \gamma=500
\end{equation}
at the stage of the interaction when plasma noticeably affects the field structure and amplitude. Thus, this estimate is also reasonable for evaluating $\gamma$ for the threshold structure.

Then, considering (\ref{Eq:targetcondition}) and Eqs. (\ref{Eq:epsilon})-(\ref{Eq:estgamma}) the relationship between the threshold initial quantity of target electrons $N_{0th}$ necessary for the formation of the threshold structure and the radius of the threshold structure are 

\begin{eqnarray}
    r_{cth}=(\sqrt{2}\pi)^{-1}\lambda\left(\left\{\left(1-\frac{n_{cth}(N_{0th},r_{cth})}{5\gamma}\right)^2+\nonumber\right.\right.\\
    \left.\left.\left(\frac{2n_{cth}(N_{0th},r_{cth})}{5\gamma}\right)^2\right\}^{0.5}-1+\frac{n_{cth}(N_{0th},r_{cth})}{5\gamma}\right)^{-0.5}\label{Eq:generalcondition},
\end{eqnarray}
where $n_{cth}(N_{0th},r_{cth})$ is given by (\ref{Eq:nc}). This equation confirms that the threshold structure is determined not solely by the threshold target density $n_{0th}$, but primarily by $N_{0th}$, in other words, by the product $n_{0th}R_0^3$.

We assume that, in the case of the threshold plasma-field structure, the real part of the dielectric constant vanishes:
\begin{equation}
    \label{Eq:conditionrealepsilon}
    \varepsilon_\mathrm{Re}\approx0.
\end{equation}
Then, from Eqs.~(\ref{Eq:targetcondition}), (\ref{Eq:epsilon}), (\ref{Eq:generalcondition}) the threshold density of plasma cylinder and its radius are 
\begin{eqnarray}
    n_{cth}=5\gamma\approx2500\label{Eq:ncval},\\
    r_{cth}=\delta=\lambda/(2\pi)\label{Eq:rcval}.
\end{eqnarray}
Since according to simulations $n_{cth}$ is around a few thousands, the above assumption is reasonable. Also note that $r_{cth}$ is very close to the displacement at which the plasma column formed due to pinching can shift from the $z$ axis and join the current sheet at such $N_0$ when the second basic structure transforms to the third one. Thus, the assumption (\ref{Eq:conditionrealepsilon}) is applicable.

As a result, the threshold quantity of electrons and the threshold density of the target as a function of target radius are
\begin{eqnarray}
    &N_{0th}\approx\frac{1.1\gamma}{f_{drag}}\frac{\lambda}{r_e}\approx1.8\times10^{12},\label{Eq:N0th}\\
    &n_{0th}=\frac{8.5\times10^{-2}\gamma}{f_{drag}}\left(\frac{\lambda}{R_0}\right)^3\approx430\left(\frac{\lambda}{R_0}\right)^3,\label{Eq:n0th}
\end{eqnarray}
where $r_e=e^2/(mc^2)\approx2.8\times10^{-13}$~cm is the classical electron radius. These estimates, relevant for spherical targets, can be generalized to other types of targets if instead of $N_{0th}$ or $n_{0th}$ we derive threshold conditions for the quantity of electrons in the plasma cylinder:
\begin{equation}
    \label{Eq:Ncth}
    N_{cth}\approx1.1\gamma\frac{\lambda}{r_e}\approx1.8\times10^{11}.
\end{equation}
If the dependence of $N_{c}$ on initial density and size of the target is known, we can obtain the initial threshold density or the threshold total quantity of electrons of the target.

Note that $N_{cth}$ is in a good agreement with the maximum quantity of pairs or electrons in focus in {\it the regime of $e^-$-$e^+$ plasma self-compression} or {\it self-compression of multicomponent plasma}. A larger quantity of particles causes a stronger field decay, reduction in the rate of pair generation and consequently a decrease in this quantity. A smaller quantity of particles ensures a strong enough field to generate pairs faster than they escape the focal region; therefore, the quantity of particles in focus should increase. So, $N_{cth}$ corresponds to a quasi-stationary quantity of particles within the first (for additional confirmation see Ref.~\cite{Efimenko_PRE2019}) or the second basic structures. For example, for a nanowire target with radius $R_{nw}\ll\lambda$ and density $n_{nw}$, we may expect that all electrons $N_{nw}=2\pi R_{nw}^2w_0 n_{nw}n_{cr}$ within part of this wire with height of $2w_0$ correspond to the electrons compressed into the focus of the dipole wave, if $N_{nw}<N_{cth}$, where $N_{cth}$ is given by Eq.~(\ref{Eq:Ncth}). So, for this type of targets the threshold quantity of electrons and initial density are $N_{nwth}\approx N_{cth}$ and $n_{nwth}\approx N_{cth}/(3R_{nw}^2\lambda n_{cr})$, respectively.

Moreover, for {\it the regime of $e^-$-$e^+$ plasma self-compression} and {\it the regime of self-compression of multicomponent plasma} we can conclude that the results of irradiation of a sphere and a nanowire are similar if the quantities $N_c$ of electrons gathered in focus due to target compression are similar too in both cases and are less than $N_{cth}$. In other words, the densities and the radii should satisfy the equation $n_0\approx7n_{nw}R_{nw}^2\lambda/R_0^3$ and the conditions $n_{nw}\lesssim N_{cth}/(3R_{nw}^2\lambda n_{cr})$ or $n_0\lesssim N_{cth}/(0.4R_0^3n_{cr})$. This interconnection of different types of targets follows from the discussion at the end of  Sec.~\ref{subsec:plasmastruct} and several examples of numerical simulations testify in favor of this conclusion (see Supplemental Material of Ref.~\cite{Efimenko_PRE2019}).

The obtained threshold values $N_{cth}$, $N_{0th}$ and $n_{0th}$ correspond to the threshold between second and third basic structures. To estimate the corresponding values of $\overline{N_{cth}}$, $\overline{N_{0th}}$ and $\overline{n_{0th}}$ related to the threshold between the first and the second basic structures we recall that these structures resemble each other: the initial target  is strongly compressed into focus, thus triggering current sheet formation and pinching. The main difference is that, for the first structure, the number of electrons compressed into focus should be much less than $N_{cth}$. In this case, the influence of electron-ion plasma on the field is negligible; vacuum breakdown is like in the given field and the quasi-neutral e$^-$-e$^+$ plasma plays the leading role. Thus, for a rough estimate we consider that
\begin{eqnarray}
    &\overline{N_{0th}}\approx0.1N_{0th}\approx1.8\times10^{11}\label{Eq:N0thl},\\
    &\overline{n_{0th}}\approx0.1n_{0th}\approx40\left(\frac{\lambda}{R_0}\right)^3\label{Eq:n0thl},\\
    &\overline{N_{cth}}\approx0.1N_{cth}\approx1.8\times10^{10}\label{Eq:Ncthl}.
\end{eqnarray}

Finally, we note that the obtained estimates of the threshold initial densities $n_{0th},~\overline{n_{0th}}$ and the initial electron quantities in the target $N_{0th},~\overline{N_{0th}}$ according to Eqs.~(\ref{Eq:n0th}), (\ref{Eq:n0thl}) and Eqs.~(\ref{Eq:N0th}), (\ref{Eq:N0thl}) are within the ranges determined in Sec.~\ref{subsec:plasmastruct}.

\subsection{\label{subsec:N}Source of e${^-}$, e${^+}$ and $\gamma$-photons}

\begin{figure*}
	\includegraphics[width=1\linewidth]{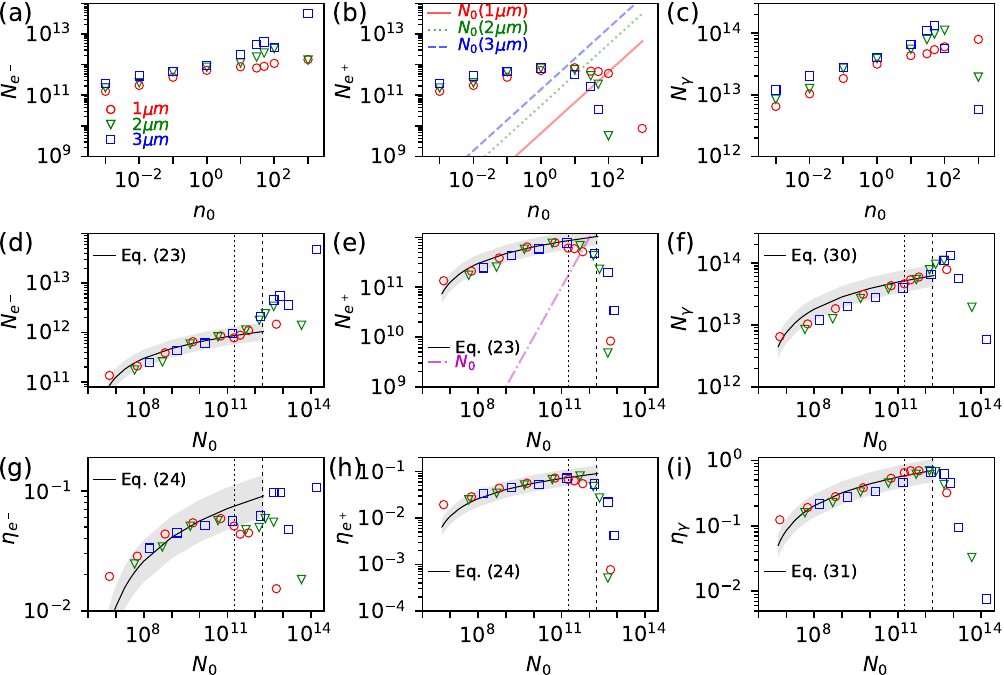}
	\caption{\label{fig:nsum}Total quantity of electrons $N_{e^-}$ (a), (d) , positrons $N_{e^+}$ (b), (e) and $\gamma$-photons $N_\gamma$ (c), (f), and efficiencies of their generation $\eta_{e^-}$ (g), $\eta_{e^+}$ (h), $\eta_\gamma$ (i), respectively, due to irradiation of spherical targets by the ideal e-dipole wave with power 30~PW and duration 15~fs as a function of initial electron density $n_0$, panels (a)-(c), and initial total electron quantity $N_0$, panels (d)-(i). Markers in all panels show results of simulations and their shapes (and colors) labeled in legends of panel (a) correspond to different  target radii. Sloping lines in (b) is the dependence of $N_0$ on $n_0$ for different target radii and in (e) the dash-dotted line shows $N_{e^+}=N_0$. Vertical dotted and dashed lines in (d)-(i) denote threshold initial electron quantities $\overline{N_{0th}}$ and $N_{0th}$ in accordance with Eqs.~(\ref{Eq:N0thl}) and (\ref{Eq:N0th}), respectively. Solid lines labeled in (d)-(i) by references to corresponding equations demonstrate approximations of numerical results. Grey color areas demonstrate the difference from estimates by no more than half.}
\end{figure*}
We start with the discussion of more rough integral characteristics of the laser-target interaction such as the quantity of accelerated electrons, positrons and $\gamma$-photons escaping the focal region, and the efficiencies of their generation as a result of target irradiation by an e-dipole wave with a power of 30~PW and a duration of 15~fs. For this simulation we accumulate information about all particles and gamma-photons escaping the focus when they cross a sphere of observation with a radius of $3.6\lambda$ and with the center at the coordinate origin. The surface of this sphere is far enough from the focus, where the main processes with particles and gamma-photons take place. Since typically in simulations the quantity of gamma-photons is approximately one-two order of magnitude larger than the quantity of electrons or positrons, we ignore the births of photons with energies less than $2mc^2$. These photons cannot generate a pair, so their absence in simulation leads only to loss of information about relatively low-energetic gamma-photons. This allows us to save memory and reduce computing time substantially.

\subsubsection{\label{subsubsec:Nest}Qualitative estimates}

Let us first generalize the simple qualitative model presented above to the quantities of the generated electrons $N_{e^-}$, positrons $N_{e^+}$ and $\gamma$-photons $N_\gamma$ as well as to the efficiencies of their generation ($\eta_{e^-,e^+,\gamma}$, respectively). The generation efficiency is calculated as a ratio of the total energy of electrons $W_{e^-}$, or positrons $W_{e^+}$, or photons $W_\gamma$ escaping the focal region and crossing the observation sphere to the total energy of laser radiation $W_L\approx 1.05P\tau\approx4.7\times10^9$~erg.

For estimates we pay attention to {\it the regime of $e^-$-$e^+$ plasma self-compression}. As discussed in Secs.~\ref{subsec:plasmastruct} and \ref{subsec:dependencetarget}, after the fast growth due to vacuum breakdown, the quantity of pairs in focus saturates at the {\it maximum} quasi-stationary level of approximately $N_{cth}$. For estimates we ignore the target compression stage and the stage of exponential growth of pair quantity (the linear stage of QED cascade, plasma influence on the fields may be neglected), since the quantity of escaping particles and photons at these stages is (much) less than the corresponding quantities at the quasi-stationary stage when the basic structure is formed. Since the height of the plasma structure along the $z$ axis is approximately $2w_0$ (see Fig.~\ref{fig:structs}(c), (g)) and each half of the wave period this structure shifts along this axis approximately by $w_0$, then the rate of pair escape $\Gamma_\mathrm{esc}$ is roughly $1/T$. A more accurate estimate based on fitting the numerical results is 
\begin{equation}
    \label{Eq:gammaesc}
    \Gamma_\mathrm{esc}=1.16/T.
\end{equation}
The duration of maintaining the first basic structure is approximately equal to the pulse duration minus the time needed to reach $N_{cth}$ from the initial quantity of electrons compressed to focus $N_c=f_{drag}N_0$ due to the QED cascade with growth rate $\Gamma$:
\begin{equation}
    \label{Eq:stationarytime}
    t_{st}=\tau-\log\left(\frac{N_{cth}}{f_{drag}N_0}\right)/\Gamma.
\end{equation}
In the case of the constant wave, $\Gamma=3.77/T$ (see Ref.~\cite{Gonoskov_PRX2017}), whereas in our case the pulse has a finite duration and $\Gamma<3.77/T$. The best fitting of numerical results gives
\begin{equation}
    \label{Eq:gammarate}
    \Gamma=2.75/T.   
\end{equation}
So, using Eqs.~(\ref{Eq:gammaesc})-(\ref{Eq:gammarate}) the estimate of the total quantity of escaping electrons and positrons as a function of $N_0$ is
\begin{equation}
    \label{Eq:Ne-e+N0}
    N_{e^-,e^+}=\Gamma_\mathrm{esc}N_{cth}\left(\tau-\log\left(\frac{N_{cth}}{f_{drag}N_0}\right)/\Gamma\right).
\end{equation}
If we assume that the electrons gain energy mainly in focus, then $\gamma$ is determined by Eq.~(\ref{Eq:estgamma}). Thus, beams of accelerated electrons and positrons are generated with the efficiency
\begin{equation}
    \label{Eq:etae-e+N0}
    \eta_{e^-,e^+}=\Gamma_\mathrm{esc}N_{cth}mc^2\gamma\left(\tau-\log\left(\frac{N_{cth}}{f_{drag}N_0}\right)/\Gamma\right)/W_L.
\end{equation}
In turn, $e^-$ and $e^+$ oscillating in focus generate photons with spectral probability density \cite{BayerKatkov}  
\begin{eqnarray}
    W'_{rad}=\frac{\alpha mc^2}{\sqrt{3}\gamma\pi\hbar}\left(\int_{y_0}^\infty{K_{5/3}(y)dy}+\right.\nonumber\\
    \left.\frac{\zeta^2}{1-\zeta}K_{2/3}(y_0)\right),\label{Eq:wraddensity}
\end{eqnarray}
where $K_\nu(y)$ is the modified Bessel function, $\nu$ is its order, $\alpha=e^2/(\hbar c)$ is the fine structure constant, $\hbar$ is the reduced Planck constant, $y_0=2\zeta/(3\chi(1-\zeta))$, $\zeta$ denotes the fraction of particle energy loss during photon emission and varies in the $(0,1)$ range.   The parameter $\chi$ is the dimensionless quantum parameter determined by the following expression \cite{Nikishov1964quantum}:
\begin{equation}
    \label{Eq:chi}
    \chi=\frac{\hbar\omega_0}{mc^2}\gamma F_\perp,
\end{equation}
where $F_\perp=\sqrt{\left(\mathbf{E}+\left[\mathbf{\beta}\times \mathbf{B}\right]\right)^2-\left(\mathbf{\beta}\cdot \mathbf{E}\right)^2}$ is the characteristic field transverse to the particle velocity $\beta$ normalized by $c$. $F_\perp$ is mainly a magnetic field around the plasma column and can be estimated as $F_\perp\approx1000$. Then, using Eq.~(\ref{Eq:estgamma}) we obtain 
\begin{equation}
    \label{Eq:chival}
    \chi\approx1.3.
\end{equation}
Thus, on the average, a photon takes away approximately the $\zeta_{av}$ part of particle energy, where
\begin{equation}
    \label{Eq:zetaav}
    \zeta_{av}=\left(\int_0^1\zeta W'_{rad}d\zeta\right)/\left(\int_0^1W'_{rad}d\zeta\right)\approx0.13,
\end{equation}
and the rate of photon emission by an electron or positron with energy higher than $2mc^2$  is 
\begin{equation}
    \label{Eq:wrad}
   W_{rad}=\int_{2/\gamma}^1 W'_{rad}d\zeta\approx35/T,
\end{equation}
By analogy with the derivation of Eqs.~({\ref{Eq:Ne-e+N0}}) and ({\ref{Eq:etae-e+N0}}), keeping in mind that $N_{cth}$ electrons and positrons in focus generate $\gamma$-photons equally, using Eq.~(\ref{Eq:wrad}) and (\ref{Eq:zetaav}) the estimate of the quantity of generated $\gamma$-photons is 
\begin{equation}
    \label{Eq:NgammaN0}
    N_\gamma\approx2W_{rad}N_{cth}\left(\tau-\log\left(\frac{N_{cth}}{f_{drag}N_0}\right)/\Gamma\right),
\end{equation}
and the estimate of the generation efficiency is
\begin{equation}
    \label{Eq:etagammaN0}
    \eta_\gamma=2W_{rad}N_{cth}mc^2\gamma\zeta_{av}\left(\tau-\log\left(\frac{N_{cth}}{f_{drag}N_0}\right)/\Gamma\right)/W_L.
\end{equation}
It is worth noting that the estimates (\ref{Eq:Ne-e+N0}), (\ref{Eq:etae-e+N0}), (\ref{Eq:NgammaN0}), (\ref{Eq:etagammaN0}) do not take into account the influence of the electron-ion plasma remaining after target compression. So, these estimates are more accurate up to $N_0\sim\overline{N_{0th}}$. Also, these estimates assume that the quantity of electrons and positrons in focus should be around $N_{cth}$ in the major part of the laser pulse (they do not take into account the linear stage of QED cascade); otherwise, for example, at lower powers or  lower $N_0$, their accuracy reduces significantly.

\subsubsection{\label{subsubsec:Numresults}Analysis of numerical results}

Armed with the simple model presented above in Sec.~\ref{subsec:dependencetarget} and \ref{subsubsec:Nest} we can verify our general understanding of the dynamics of laser-target irradiation and specify quantitative outcomes of this interaction based on the numerical results shown in Fig.~\ref{fig:nsum}. First, the comparison of the results presented as a function of $n_0$ (Figs.~\ref{fig:nsum}(a)-(c)) and $N_0$ (Figs.~\ref{fig:nsum}(d)-(i)) confirms the conclusion that in the zeroth approximation $N_0$ is the similarity parameter for {\it the regime of $e^-$-$e^+$ plasma self-compression} and {\it the self-compression of multicomponent plasma} ($N_0<N_{0th}$) resulting in the  formation of the first and second basic structures. Despite the comparable behavior of the quantities of escaping positrons $N_{e^+}$ and photons $N_\gamma$ as well as the efficiencies of their generation ($\eta_{e^+}$, $\eta_\gamma$, respectively) even at $N_0>N_{0th}$ at different target radii (Figs.~\ref{fig:nsum}(e), (f) and (h), (i)), functions $N_{e^-}(N_0>N_{0th})$ as well as $\eta_{e^-}(N_0>N_{0th})$ (Figs.~\ref{fig:nsum}(d), (g)) are much less correlated.  The loss of correlations of these values at $N_0>N_{0th}$ also confirms our estimate of the threshold initial electron quantity of the third basic structure in Eq.~(\ref{Eq:N0th}) or, consequently, the threshold initial electron density in Eq.~(\ref{Eq:n0th}) and that there are different variations of this structure. 

Second, such  target irradiation  by laser pulses in the form of the e-dipole wave can be a very effective source of $\gamma$-photons. The maximum quantity of $\gamma$-photons $N_\gamma^\mathrm{max}\approx10^{14}$ is at $N_0\approx5\times10^{12}$ and the maximum efficiency $\eta_\gamma^\mathrm{max}\approx70\%$ is at $N_0\approx2\times10^{12}$. So, the maxima of these values are reached approximately at $N_0\approx N_{0th}$ with an accuracy up to a coefficient of the order of unity, because when the third basic structure is formed, the field penetration inside plasma is strongly suppressed. Thus, the particles can gain less energy, $\chi\ll1$; in this case, $\zeta_{av}\sim\chi$, $W_{rad}\sim F_\perp$ \cite{BayerKatkov}, so the efficiency of the generation of $\gamma$-photons as well as their quantity decreases at $N_0\gtrsim N_{0th}$ (Figs.~\ref{fig:nsum}(f), (i)). A consequence of this is an even sharper decay of $N_{e^+}$ and $\eta_{e^+}$ within this range of $N_0$ (Figs.~\ref{fig:nsum}(e), (h)).

Third, positron generation can be rather effective too but at other target parameters. The maximum generation of positrons will be when the quasi-stationary basic structure with a maximum quantity of particles in focus is maintained for a longer period of time and the initial target plasma has minimum effects. Thus, this maximum can be reached at the boundary between the first and the second basic structures. Figures~\ref{fig:nsum}(e) and (h) confirm that our estimate of $\overline{N_{0th}}$ (\ref{Eq:N0thl}) corresponds well to this threshold. According to the obtained numerical results, the maximum quantity of generated positrons $N_{e^+}^\mathrm{max}\approx8\times10^{11}$ (total charge around 100~nC) and $\eta_{e^+}^\mathrm{max}\approx8\%$. Since $N_{e^+}^\mathrm{max}$ is reached at the maximum applicability limit of the derived estimates ($N_0\sim\overline{N_{0th}}$ or $N_{cth}/(f_{drag}N_0)\sim0.1$), we can also estimate $N_{e^+}^\mathrm{max}$ using Eq.~(\ref{Eq:Ne-e+N0}) as
\begin{equation}
    \label{Eq:Nposmax}
    N_{e^+}^\mathrm{max}\approx\Gamma_\mathrm{esc}N_{cth}\left(\tau-2.3/\Gamma\right),
\end{equation}
and the maximum efficiency of their generation using Eq.~(\ref{Eq:etae-e+N0}) as
\begin{equation}
    \label{Eq:etaposmax}
    \eta_{e^+}^\mathrm{max}\approx\Gamma_\mathrm{esc}N_{cth}mc^2\gamma\left(\tau-2.3/\Gamma\right)/W_L.
\end{equation}
For the considered e-dipole wave, these estimates give $N_{e^+}^\mathrm{max}\approx8.7\times10^{11}$ and $\eta_{e^+}^\mathrm{max}\approx7.6\%$, which are very close to the numerical results. Note that at such target parameters the quantity of escaping electron and the efficiency of their generation are quite large too, because in this regime the quantities of electrons and positrons are comparable and $e^-$-$e^+$ along with electron-ion plasma significantly influence the dynamics of interaction with laser radiation. 

\begin{figure*}
	\includegraphics[width=1\linewidth]{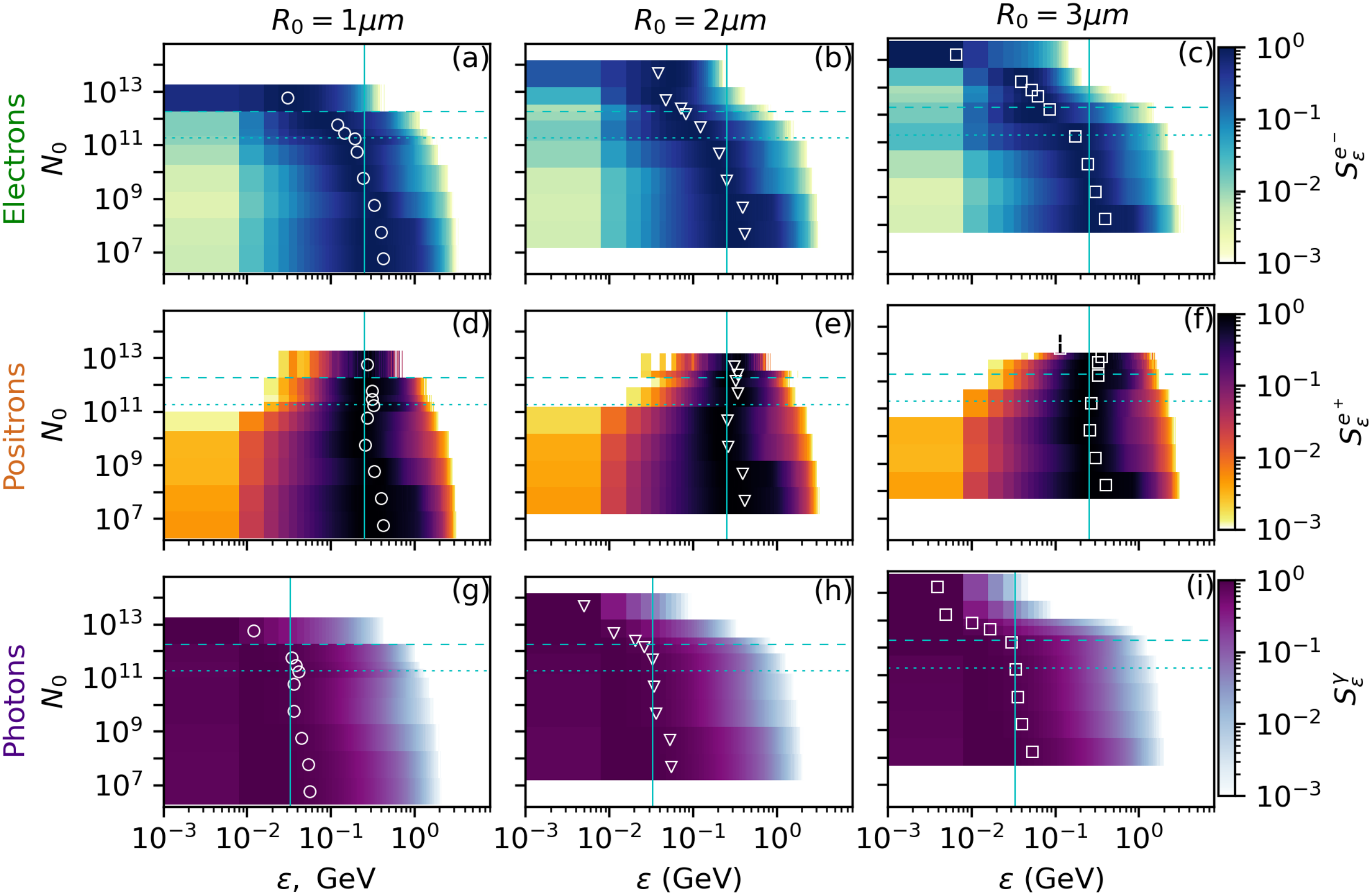}
	\caption{\label{fig:ensp} Energy spectra of $e^-$ ($S^{e^-}_\varepsilon$) (a)-(c), $e^+$ ($S^{e^+}_\varepsilon$) (d)-(f) and $\gamma$-photons ($S^\gamma_\varepsilon$) (g)-(i) generated as a result of  irradiation of a target with radius $R_0=1~\mu$m (a), (d), (g), $R_0=2~\mu$m (b), (e), (h), $R_0=3~\mu$m (c), (f), (i) by the e-dipole wave with power of 30~PW and duration of 15~fs. Horizontal dotted and dashed lines show $\overline{N_{0th}}$ (Eq.~(\ref{Eq:N0thl})) and $N_{0th}$ (Eq.~(\ref{Eq:N0th})), respectively.  Markers denote the average energy of particles and photons retrieved from numerical results. Vertical solid lines correspond to estimates of average energies of particles and photons according to Eqs.~(\ref{Eq:estgamma}) and (\ref{Eq:zetaav}).}
\end{figure*}

Fourth, since the electrons are generated not only as a result of the vacuum breakdown but also due to the ionization of the initial target, the quantity of escaping accelerated electrons and the efficiency of their generation can grow even when the third basic plasma structure is formed and the generation of positrons and $\gamma$-photons decreases. The behavior of $N_{e^-}$ and $\eta_{e^-}$ at $N_0>N_{0th}$ is quite complex and non-monotonic and is not determined only by $N_0$ or $n_0$. This reflects the formation of different variations of the third basic structure, which should be studied in more detail depending on target size, density, and shape for determining the most efficient ones. Nevertheless, the performed numerical simulations show that the maximum quantity of escaping electrons $N_{e^-}^\mathrm{max}\approx4.8\times10^{13}$ (around 7~$\mu$C) and the maximum efficiency $\eta_{e^-}^\mathrm{max}\approx10\%$ are reached for the target with radius $R_0=3\mu$m and initial density $n_0=1000$, that is the target with maximum considered $N_0$. Since $\eta_{e^+}^\mathrm{max}$ is close to $\eta_{e^-}^\mathrm{max}$ and the difference between $N_{e^-}^\mathrm{max}$ and $N_{e^+}^\mathrm{max}$ is more than an order of magnitude, the majority of electrons are accelerated to energies an order of magnitude lower than positron energies.

Finally, note that the derived functions $N_{e^-,e^+}$ (\ref{Eq:Ne-e+N0}), $\eta_{e^-,e^+}$ (\ref{Eq:etae-e+N0}) and $N_\gamma$ (\ref{Eq:NgammaN0}), and $\eta_\gamma$ (\ref{Eq:etagammaN0}) demonstrate the dependence on $N_0$ that is very similar to the behavior of the functions retrieved from numerical data (see Figs.~\ref{fig:nsum}(d)-(i)). In general, the difference of estimates and numerical values is not larger than 50\%, which is rather good for qualitative estimates. The exception in the $N_0<10^8$ range is caused by ignoring for estimates the linear stage of the QED cascade, while maintenance of the quasi-stationary basic structure is too short within this range. As a result, the obtained functions lead to underestimation of the quantities of particles and photons and the efficiencies of their generation within this range. At the same time, these functions give reasonable approximations even above the region where they are assumed to be valid, not only at $N_0<\overline{N_{0th}}$ (in {\it the regime of $e^-$-$e^+$ plasma self-compression}) but also at $\overline{N_{0th}}<N_0<N_{0th}$ (in {\it the regime of self-compression of multicomponent plasma}). Since for estimates we suppose equal quantities of electrons and positrons, the electron quantity $N_{e^-}$, being larger in the second regime, is underestimated in (\ref{Eq:Ne-e+N0}) while positron quantity $N_{e^+}$ is overestimated (see Figs.~\ref{fig:nsum}(d) and (e)).

\subsubsection{Energy spectra and angular distributions}

\begin{figure*}
	\includegraphics[width=1\linewidth]{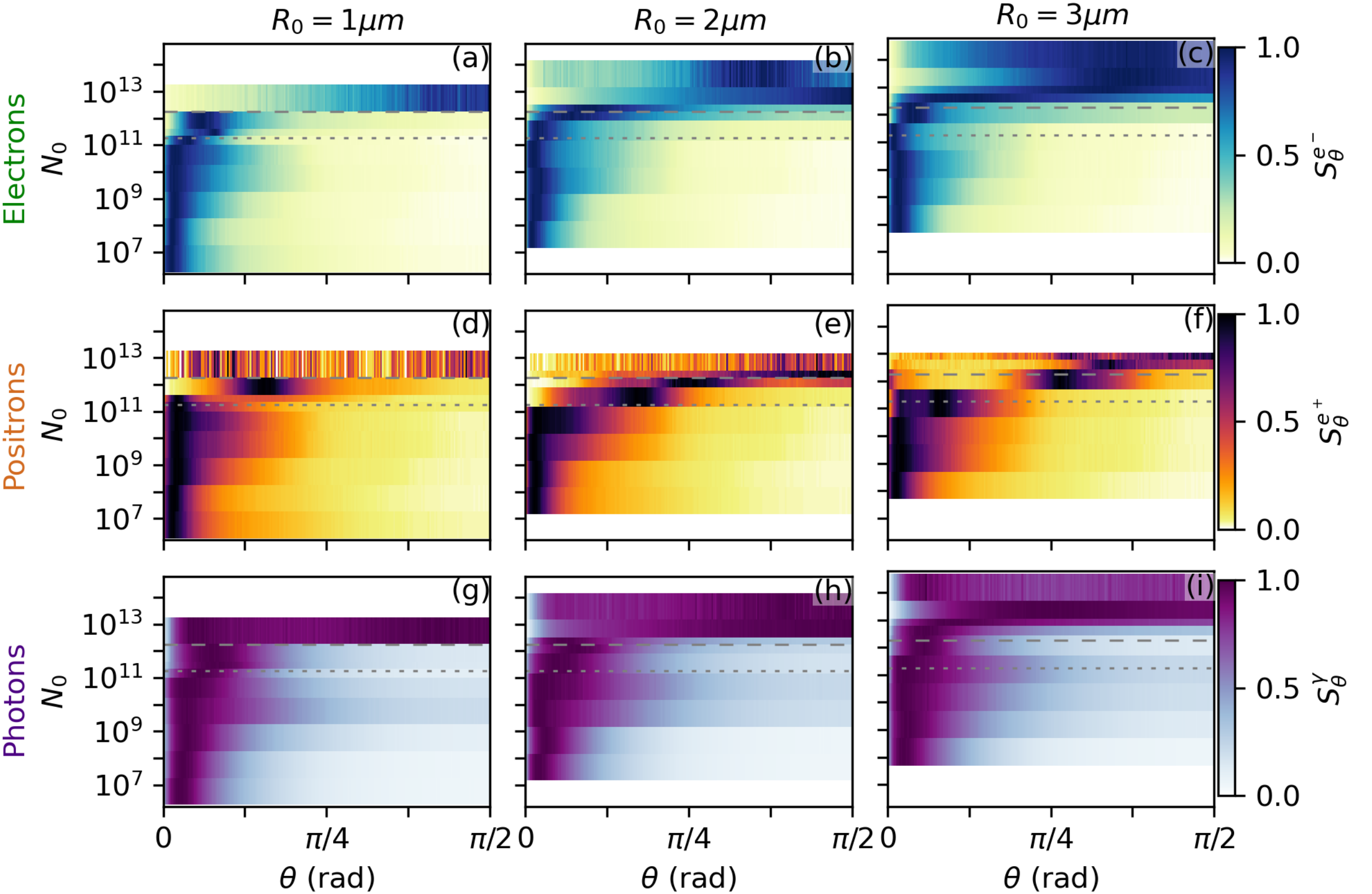}
	\caption{\label{fig:theta} Angular distributions of $e^-$ ($S^{e^-}_\theta$) (a)-(c), $e^+$ ($S^{e^+}_\theta$) (d)-(f) and $\gamma$-photons ($S^\gamma_\theta$) (g)-(i) generated as a result of irradiation of target with radius $R_0=1~\mu$m (a), (d), (g), $R_0=2~\mu$m (b), (e), (h), $R_0=3~\mu$m (c), (f), (i) by the e-dipole wave with 30~PW power and 15~fs duration .}
\end{figure*}

Besides the quantity of particles and $\gamma$-photons, their spectra and angular distributions are crucially important in practice. In simulations we retrieve these characteristics based on the particles and photons escaping the focus. We record energy $\varepsilon$ of all particles and photons crossing the observation sphere and obtain the energy distribution $\partial W_{e^-,e^+\gamma}/\partial\varepsilon$ after each simulation. Then for each target radius and various $n_0$ or $N_0$ we stack up these distributions normalized by their maxima $\left(\partial W_{e^-,e^+\gamma}/\partial\varepsilon\right)/\left(\partial W_{e^-,e^+\gamma}/\partial\varepsilon\right)|_\mathrm{max}$ and finally obtain a spectral map $S_\varepsilon^{e^-,e^+,\gamma}$ as a function of $N_0$, $R_0$ and $\varepsilon$.

The retrieval of the angular distribution is very similar. Initially, for particles and photons we record the angular distribution $\partial W_{e^-,e^+\gamma}/\partial\theta$ of the total energy of the particles and photons crossing the observation sphere throughout the simulation, where $\theta=\arccos{(p_z/p)}$ is the polar angle, $p$ is the magnitude of momentum of the particles and photons, and $p_z$ is its $z$ component. Then we stack up  the distributions $\partial W_{e^-,e^+\gamma}/\partial\theta$ normalized by their maxima and obtain a map of angular distribution $S^{e^-,e^+,\gamma}_\theta$ as a function of $N_0$, $R_0$, $\theta$.

Note that the angular distribution in the far field (at the distance much larger than the wavelength) is of practical interest. In order to retrieve a far field distribution, it is assumed that the particles and photons crossing the observation sphere propagate from the coordinate origin and the perturbations of particle trajectories (trajectories of the photons are straight lines) outside this sphere due to weaker fields may be neglected. Thus, the distributions over $\theta$ calculated for particles within the simulation box in momentum space correspond to the distribution over $\theta$ calculated in the far field in coordinate space. Also, keeping in mind a high axial symmetry of the fields of the dipole configuration we consider the angular distribution characterizing the energy emitted into a cone layer (rather than into the element of the solid angle) characterized by angle $\theta$. In other words, we consider the distribution characterizing the energy emitted into the space between two cones with the $z$ axis as the symmetry axis and with opening angles $2\theta$ and $2(\theta+d\theta)$ when $d\theta\rightarrow0$.

The obtained numerical results shown in Figs.~\ref{fig:ensp} and \ref{fig:theta} demonstrate good correlation with the previously discussed plasma-field dynamics. These figures confirm that $N_0$ is the similarity parameter for both energy spectra and angular distributions because qualitative changes in $S^{e^-,e^+,\gamma}_\varepsilon$ and $S^{e^-,e^+,\gamma}_\theta$ occur at approximately threshold quantities $N_{0th}$ and $\overline{N_{0th}}$. So let us discuss the spectra and angular distributions corresponding to different regimes of laser-target interaction within different ranges of $N_0$. 

At relatively low $N_0$ ($N_0<\overline{N_{0th}}$) in {\it the regime of $e^-$-$e^+$ plasma self-compression}, the energies of particles and $\gamma$-photons are highest, because the fields penetrate throughout the plasma distribution and the influence of the initial electron-ion plasma on the fields is negligible. Particle and photon energies reach a maximum of a few GeV with approximately the same cut-off. However, highly energetic particles and photons represent a small part of all particles and photons. The average energy of particles is 250-400~MeV (Figs.~\ref{fig:ensp}(a)-(f)) and the average energy of $\gamma$-photons is an order of magnitude less, approximately 30-50~MeV (Figs.~\ref{fig:ensp}(g)-(i)). These values agree with an accuracy up to a coefficient of the order of unity with estimates $mc^2\gamma\approx250$~MeV and $\zeta_{av}mc^2\gamma\approx30$~MeV, respectively, based on Eqs.~(\ref{Eq:estgamma}) and (\ref{Eq:zetaav}). While the average energy of positrons and photons is quasi-constant, the average electron energy slowly monotonically decreases. This occurs because, besides more energetic electrons generated in the region of the strongest fields due to the QED cascade, at $N_0$ close to or larger than $\overline{N_{0th}}$, there is a noticeable quantity of less energetic electrons which are originated due to target ionization and are scattered on their way to the focus without gain in high energy. 

Also, in the first regime, the directivity of the particles and $\gamma$-photons  escape from the focus is optimal  (Fig.~\ref{fig:theta}). The particles and photons propagate in a cone layer at $\theta\approx3^\circ$ and $\theta\approx7^\circ$, respectively. The angular width of this layer increases with an increase in $N_0$ from $10^\circ$ to $20^\circ$ for particles and from $15^\circ$ to $30^\circ$ for photons.

At moderate $N_0$ ($\overline{N_{0th}}<N_0<N_{0th}$) in {\it the regime of self-compression of multicomponent plasma}, the initial electron-ion plasma starts to influence the laser plasma-interaction. As a result, maximum energies of particles and photons decrease down to about 1~GeV at $N_0=N_{0th}$. However, since the fields penetrate well into plasma, the average energy of positrons and photons remains approximately the same (Fig.~\ref{fig:ensp}(d)-(f) and Fig.~\ref{fig:ensp}(g)-(i)). An exception is the decay of the average electron energy for the same reason as in the first regime (Fig.~\ref{fig:ensp}(a)-(c)). Because of a wider plasma-field structure in the second regime (compare Figs.~\ref{fig:structs}(a)-(d) and Figs.~\ref{fig:structs}(e)-(f)), the particles escaping from the focus can be deflected at a larger polar angle and the maximum of $S_\theta^{e^+}$ is at $\theta$ from $5^\circ$ to $45^\circ$ (Fig.~\ref{fig:theta}(d)-(f)). In turn, the photons propagate at the angles approximately equal to twice the angles of photon propagation in the first regime (Fig.~\ref{fig:theta}(g)-(i)). 

At high $N_0$ ($N_0>N_{0th}$) in {\it the regime of weak laser compression of the target} due to a strong suppression of field penetration into plasma, the maximum and average energies decrease down to 200~MeV and 10~MeV for electrons (Fig.~\ref{fig:ensp}(a)-(c)) and down to 30~MeV and 3~MeV for photons (Fig.~\ref{fig:ensp}(g)-(i)), respectively. As a result of this decrease the quantity of positrons reduces and their distributions become striped and noisy at $N_0\sim N_{0th}$ and at $N_0\gg N_{0th}$ positrons disappear (Fig.~\ref{fig:theta}(d)-(f)). Since the size of the plasma-field distribution becomes comparable with or even larger than the wavelength, the shape of the basic plasma-field structure becomes oval rather than cylindrical. So, electrons and photons can escape the focus at large $\theta$ and angular distributions become almost quasi-uniform (Fig.~\ref{fig:theta}(a)-(c) and Fig.~\ref{fig:theta}(g)-(i)).

Thus, each regime of laser-target interaction is characterized by certain spectral and angular distributions, and we can control these characteristics by choosing appropriate $N_0$.

\subsubsection{Particles and photons of GeV energies}

\begin{figure*}
	\includegraphics[width=1\linewidth]{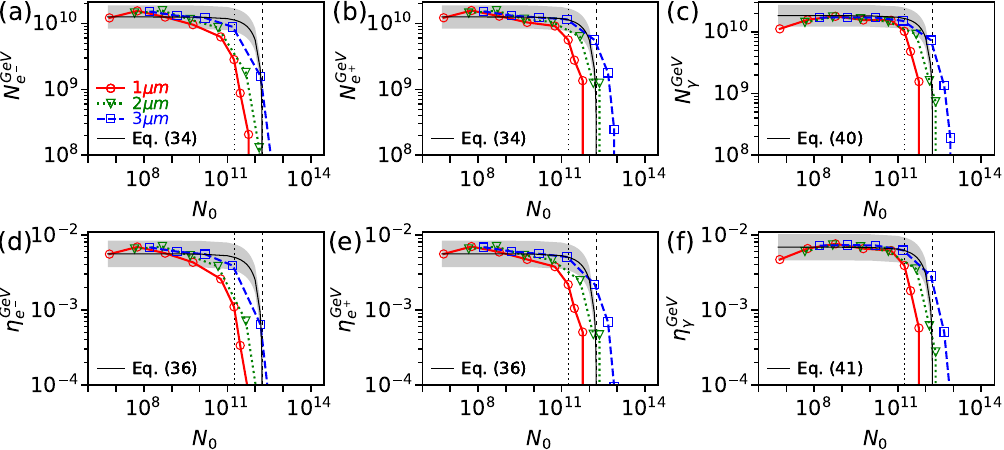}
	\caption{\label{fig:ngev} Total quantity of electrons $N_{e^-}^{GeV}$ (a), positrons $N_{e^+}^{GeV}$ (b) and $\gamma$-photons $N_\gamma^{GeV}$ (c) with energies higher than 1~GeV, and efficiencies of their generation $\eta_{e^-}^{GeV}$ (d), $\eta_{e^+}^{GeV}$ (e), $\eta_\gamma^{GeV}$ (f), respectively, due to irradiation of spherical targets by ideal e-dipole wave with 30~PW power and 15~fs duration as a function of initial total quantity $N_0$ of electrons. Markers in all panels show results of simulations and their shapes (and colors) labeled in legends of panel (a) correspond to different target radii. Vertical dotted and dashed lines denote threshold initial electron quantities $\overline{N_{0th}}$ and $N_{0th}$ in accordance with Eqs.~(\ref{Eq:N0thl}) and (\ref{Eq:N0th}), respectively. Grey color areas demonstrate the difference from estimates by no more than half.}
\end{figure*}

As shown in the previous section, the spectra of $e^-$, $e^+$ and $\gamma$-photons are quite wide and extend to the GeV range. This range is of particular interest especially for nuclear physics \cite{ELI-NP,Spring-8,A2,Nedorezov_PU2004}. Moreover, the laser-target interaction can be a source of such particles and photons, even with record fluxes \cite{Zhu_NC2016,Luo_PPCF2018,Gonoskov_PRX2017,Magnusson_PRL2019} in comparison with conventional sources like synchrotrons and Compton backscattering sources \cite{Spring-8,Nedorezov_PU2004}.

Within the considered laser-target interaction, particles and photons with GeV energies correspond to the tails of the spectra, they almost do not affect the plasma-field dynamics, but the generation of these highly energetic particles and photons is very sensitive to the field strengths. In {\it the regime of weak laser compression of the target}, field amplitudes are (much) weaker than 1000 (Fig.~\ref{fig:structs}(i), (j), (m), (n)), and the generation of GeV particles and photons is suppressed. The results of simulation (Fig.~\ref{fig:ngev}) confirm that the quantity of $e^-$, $e^+$ and photons with GeV energies ($N_{e^-}^{GeV}$, $N_{e^+}^{GeV}$, $N_{\gamma}^{GeV}$, respectively) as well as the efficiency of their generation ($\eta_{e^-}^{GeV}$, $\eta_{e^+}^{GeV}$, $\eta_{\gamma}^{GeV}$, respectively) vanish at $N_0\gtrsim N_{0th}$. In {\it the regimes of $e^-$-$e^+$ and multicomponent plasma self-compression} at the stages when the first and the second basic structures are formed, maximum field amplitudes are around 1000 (Fig.~\ref{fig:structs}(a), (b), (e), (f)) and maximum energies are around $1000mc^2<1$~GeV. So, only at the linear stage of QED cascade (when plasma influence on the field may be neglected) in both these regimes (at $N_0<N_{0th}$) we can expect efficient generation of GeV particles and photons. This agrees with the numerical results (Fig.~\ref{fig:ngev}).

Maximum outputs of GeV particles and photons occur if the quantity of pairs in focus approaches the threshold quantity $N_{cth}$ (see Eq.~(\ref{Eq:Ncth})), when the laser pulse peak reaches the focus. This was shown for photons in Ref.~\cite{Gonoskov_PRX2017}, and our results confirm the validity of this condition for pairs. Optimum target parameters correspond to {\it the regime of $e^-$-$e^+$ plasma self-compression} ($N_0<\overline{N_{0th}}$). The maximum quantities of GeV particles and photons of around $1.6\times10^{10}$ (around 2~nC for particles) and the efficiencies of their generation of around 0.7\% are achieved when $N_0\approx5\times10^8$.

Note that the quantities and efficiencies are not less than a half of optimum within a wide range of target parameters, within $10^7\lesssim N_0\lesssim\overline{N_{0th}}$ (Fig.~\ref{fig:ngev}). This range is quite wide because the largest portion of pairs is generated when the field amplitudes in focus are close to their maximum and the quantity of pairs in focus approaches $N_{cth}$ nearly over the same period of time. Thus, these results are quite robust and any reasonable mismatch of laser and target parameters does not get us far from optima. If $N_0<10^8$, then the quantity of pairs in focus approaches the threshold quantity $N_{cth}$ due to the QED cascade only at the end of the laser pulse. If $N_0\gtrsim\overline{N_{0th}}$, then the plasma initially gathered in focus due to target compression begins to weaken the fields and the quantities of GeV particles and photons decreases.

Also, a very good coincidence of $N_{e^-,e^+,\gamma}^{GeV}$ and $\eta_{e^-,e^+,\gamma}^{GeV}$ as a function of $N_0$ for different target radii emphasizes that $N_0$ is the similarity parameter in the zeroth approximation, while $N_0\lesssim\overline{N_{0th}}$. Any deviation from a unified function is mainly caused by finite pulse duration. At $N_0\gtrsim \overline{N_{0th}}$ values of $N_{e^-,e^+,\gamma}^{GeV}$ and $\eta_{e^-,e^+,\gamma}^{GeV}$ are determined not only by $N_0$, but also by $R_0$. 

The obtained numerical results enable simple qualitative estimation of $N_{e^-,e^+,\gamma}^{GeV}$ and $\eta_{e^-,e^+,\gamma}^{GeV}$. Since the range of optimal target parameters is quite wide, we use the constant wave approach for derivation. This means that the quantity of pairs in focus always reaches $N_{cth}$ when $N_0<N_{0th}$. Based on the above analysis we assume that the e-dipole wave gathers approximately $f_{drag}N_0$ electrons in focus due to the initial laser compression of the target and the generation of GeV particles and photons continues, while the quantity of pairs in focus is $f_{drag}N_0\exp{(\Gamma t)}<N_{cth}$, where $f_{drag}\approx0.1$. The pairs escape from the focus with the rate $\Gamma_{esc}$. Thus, the total quantity of escaping GeV particles is 
\begin{equation}
	\label{Eq:Neesc}
	N_{e^-,e^+}^{GeV}=f_{GeV}\Gamma_{esc}\left(N_{cth}-f_{drag}N_0\right)/\Gamma,
\end{equation}
where $f_{GeV}$ (much less than unity, $\approx0.16$ according to the simulations) is  part of the particles having energies higher than 1~GeV.

At the linear stage of the cascade, when the wave field is like in the vacuum, the particle energy will be $\approx mc^2E_a$ \cite{Gonoskov_PRL2014}. According to the simulation, it will be on the average 
\begin{equation}
    \label{Eq:gamgev}
    \gamma_{GeV}\approx0.6mc^2E_a,
\end{equation}
where $E_a$ is defined by Eq.~(\ref{Eq:E_P}). Note that this energy is higher than the energy (Eq.~(\ref{Eq:estgamma})) when basic structures are formed. Thus, the efficiency of generation of pairs and their acceleration up to GeV energies is 
\begin{equation}
	\label{Eq:etaeesc}
	\eta_{e^-,e^+}^{GeV}=f_{GeV}\Gamma_{esc}\gamma_{GeV}mc^2\left(N_{cth}-f_{drag}N_0\right)/(\Gamma W_L).
\end{equation}
With such $f_{GeV}$ estimates (\ref{Eq:Neesc}) and (\ref{Eq:etaeesc})  agree well with the numerical results (Fig.~\ref{fig:ngev}(a), (b), (d) and (e)) at $N_0<\overline{N_{0th}}$.

The quantity of GeV photons and the efficiencies of their generation can be estimated in a similar way. Their quantity is determined by the total quantity of the pairs accelerated in focus and the rate of emission of a GeV photon $W_{rad}^{GeV}$ by a particle with energy $\gamma_{GeV}mc^2$. The parameter $\chi$ at the linear stage is determined by the amplitude of the magnetic field in vacuum $B_a$ with coefficient 0.2 according to the simulations
\begin{equation}
    \label{Eq:chigev}
    \chi^{GeV}\approx0.2\frac{\hbar\omega_0}{mc^2}\gamma_{GeV}B_a\approx3.9,
\end{equation}
where $B_a$ is from Eq.~(\ref{Eq:B_P}). Note that this parameter is larger than $\chi$ at the stage when basic structures are formed. Then $W_{rad}^{GeV}$ can be obtained from Eq.~{\ref{Eq:wraddensity}} using Eqs.~(\ref{Eq:gamgev}), (\ref{Eq:chigev}) and the low limit of the integral $2000/\gamma_{GeV}$:
\begin{equation}
    \label{Eq:wradgev}
    W_{rad}^{GeV}=0.25\int_{2000/\gamma_{GeV}}^1 W_{rad}'d\zeta\approx0.14/T,
\end{equation}
where the coefficient 0.25 is due to the fact that photons with the highest energy can be emitted equally by electrons and positrons during $T/8$ twice per period \cite{Gonoskov_PRX2017}. The average fraction of the energy lost by GeV particles as a result of photon emission is:
\begin{equation}
    \label{Eq:zetagev}
    \zeta^{Gev}_{av}=\left(\int_{2000/\gamma_{GeV}}^1\zeta W'_{rad}d\zeta\right)/\left(\int_{2000/\gamma_{GeV}}^1W'_{rad}d\zeta\right)\approx0.85.
\end{equation}
Thus, the total quantity of GeV photon and the efficiency of their generation are
\begin{eqnarray}
    &N_{\gamma}^{GeV}\approx 2W_{rad}^{GeV}\left(N_{cth}-f_{drag}N_0\right)/\Gamma,\label{Eq:Nphgev}\\
    &\eta_{\gamma}^{GeV}\approx 2W_{rad}^{GeV}mc^2\gamma_{GeV}\zeta_{av}^{GeV}\left(N_{cth}-f_{drag}N_0\right)/(\Gamma W_L).\label{Eq:etaphgev}
\end{eqnarray}
Equations~(\ref{Eq:Nphgev}) and (\ref{Eq:etaphgev}) give reasonable estimates (Fig.~\ref{fig:ngev}(c) and (f)) and confirm that the results are robust because within a wide range of target parameters the generation of GeV photons is close to optimal. 

Note that the estimates (\ref{Eq:Nphgev}) and (\ref{Eq:etaphgev}) like  (\ref{Eq:Neesc}) and (\ref{Eq:etaeesc}) are more accurate at $N_0\lesssim\overline{N_{0th}}$, because they do not include the impact of electron-ion plasma on the laser-plasma dynamics. But their inaccuracy increases significantly for the parameters at which only the linear regime of the QED cascade is realized.

\subsection{Breakdown threshold}
There is one more consequence based on the determined quantities of escaping particles and photons. As is clear from Fig.~\ref{fig:nsum}(b) or (e), for some target parameters ($N_0>N_{0th}$) it is hard to argue that the vacuum breakdown occurs (too few positrons in comparison with $N_0$ are generated), though the wave power is higher than the threshold power (about 7~PW \cite{Gonoskov_PRX2017,Efimenko_SR2018}) of the vacuum breakdown, which implies abundant pair production. An explanation of this paradox is in the definitions of the vacuum breakdown thresholds. The first possible definition claims that the breakdown occurs when particle losses (mainly because of the escape from the strong field region due to field inhomogeneity) are balanced by the rate of particle generation (as a result of the QED cascade). Using this definition for a constant wave under the conditions when the plasma influence on the fields is negligible, the threshold power of about 7~PW  was obtained \cite{Gonoskov_PRX2017,Efimenko_SR2018}. This definition is more fundamental and the threshold is caused by particle motion, photon emission and photon decay in a certain field configuration. For the case of the finite laser pulse  considered here, this definition is also reasonable, because according to it while the instantaneous power is higher than 7~PW and the fields generated by plasma are much weaker than the laser fields, the quantity of pairs in focus grows with the rate (determined in Ref.~\cite{Gonoskov_PRX2017}) corresponding to the instantaneous power. This is especially relevant for the formation of the first basic structure.

However, from the practical point of view, another definition is possible, taking into account finite duration of laser pulses and their interaction with a real target. The target can be completely destroyed and does not give any seed particles when the instantaneous power exceeds 7~PW or can significantly suppress field penetration into plasma, so that photon decay becomes improbable in such fields. Thus, the second definition used in a number of papers \cite{Gelfer_PRA2015,Jirka_SR2017,Tamburini_SR2017} assumes that the threshold is reached if the quantity of the generated $e^-$-$e^+$ pairs is at least equal to the initial quantity of electrons $N_0$ in the target.

We believe that, being more fundamental, the first definition allows determining mainly necessary conditions, while the second definition determines whether the necessary conditions are sufficient in practice. So, using the first definition we can determine the required power, while the second definition answers us whether laser pulse duration and target can ensure a sufficient quantity of generated pairs.

From the performed numerical simulations (see Fig.~\ref{fig:nsum}(e) or (b)) it is clear that we can determine only maximum threshold initial density $n_{vb}^\mathrm{max}$ (or maximum threshold initial quantity of electrons in the target $N_{vb}^\mathrm{max}$) above which the plasma of the target prevents field penetration and vacuum breakdown is impossible. According to the simulations, the threshold target parameters for vacuum breakdown are $n_{vb}^\mathrm{max}\approx90,~10,~4$ (or $N_{vb}^\mathrm{max}\approx5.3\times10^{11},~5\times10^{11},~6.7\times10^{11}$) for targets with radii $R_0=1,~2,~3~\mu$m, respectively. It is to be expected that the threshold maximum density or the threshold maximum initial quantity of electrons corresponds to the second basic structure which assumes reduction of the generated quantity of positrons. We can roughly consider that $N_{vb}^\mathrm{max}\approx0.5N_{0th}\approx6\times10^{11}$ and $n_{vb}^\mathrm{max}\sim R_0^{-3}$. This allows simply choosing an appropriate target for the breakdown by varying material and target size. Also we can confirm that in the zeroth approximation $N_0$ is the similarity parameter also relevant for vacuum breakdown.

In our simulation we do not consider low densities at which the vacuum breakdown is impossible. However, it is clear that there exists a minimum threshold initial density $n_{vb}^\mathrm{min}$ (or a minimum threshold initial quantity of electrons in the target $N_{vb}^\mathrm{min}$). At this density, the initial target is completely destroyed by the leading edge of the laser pulse due to field inhomogeneity. Nevertheless, we can use the estimate (\ref{Eq:Ne-e+N0}) in order to determine both thresholds from the equation
\begin{equation}
    \label{Eq:breakcondmin}
    \Gamma_{esc}N_{cth}\left(\tau-\log\left(\frac{N_{cth}}{f_{drag}N_{vb}}\right)/\Gamma\right)=N_{vb}
\end{equation}
Equation~(\ref{Eq:breakcondmin}) has two roots. The first root is $N_{vb}^\mathrm{max}\approx10^{12}$ and is approximately 1.7 times larger than $N_{vb}^\mathrm{max}$ retrieved from simulations. The reason for this overestimation is discussed in Sec.~\ref{subsubsec:Numresults}. Nevertheless, the estimate demonstrates a good agreement with the numerical results. The second root $N_{vb}^\mathrm{min}\approx2\times10^6$ also gives the upper estimate, since Eq.~(\ref{Eq:Ne-e+N0}) underestimates $N_{e^+}$ at $N_0<10^8$. According to Eq.~(\ref{Eq:n0N0}) $n_{vb}^\mathrm{min}\approx3\times10^{-4}/R_{\mu m}^3$ ($5\times10^{17}/R_{\mu m}^3$~cm$^{-3}$ in dimensional units), so the minimum threshold density for the breakdown corresponds to gas or even residual gas as an initial target. The minimum threshold should be determined more rigorously.

\begin{figure*}
	\includegraphics[width=1\linewidth]{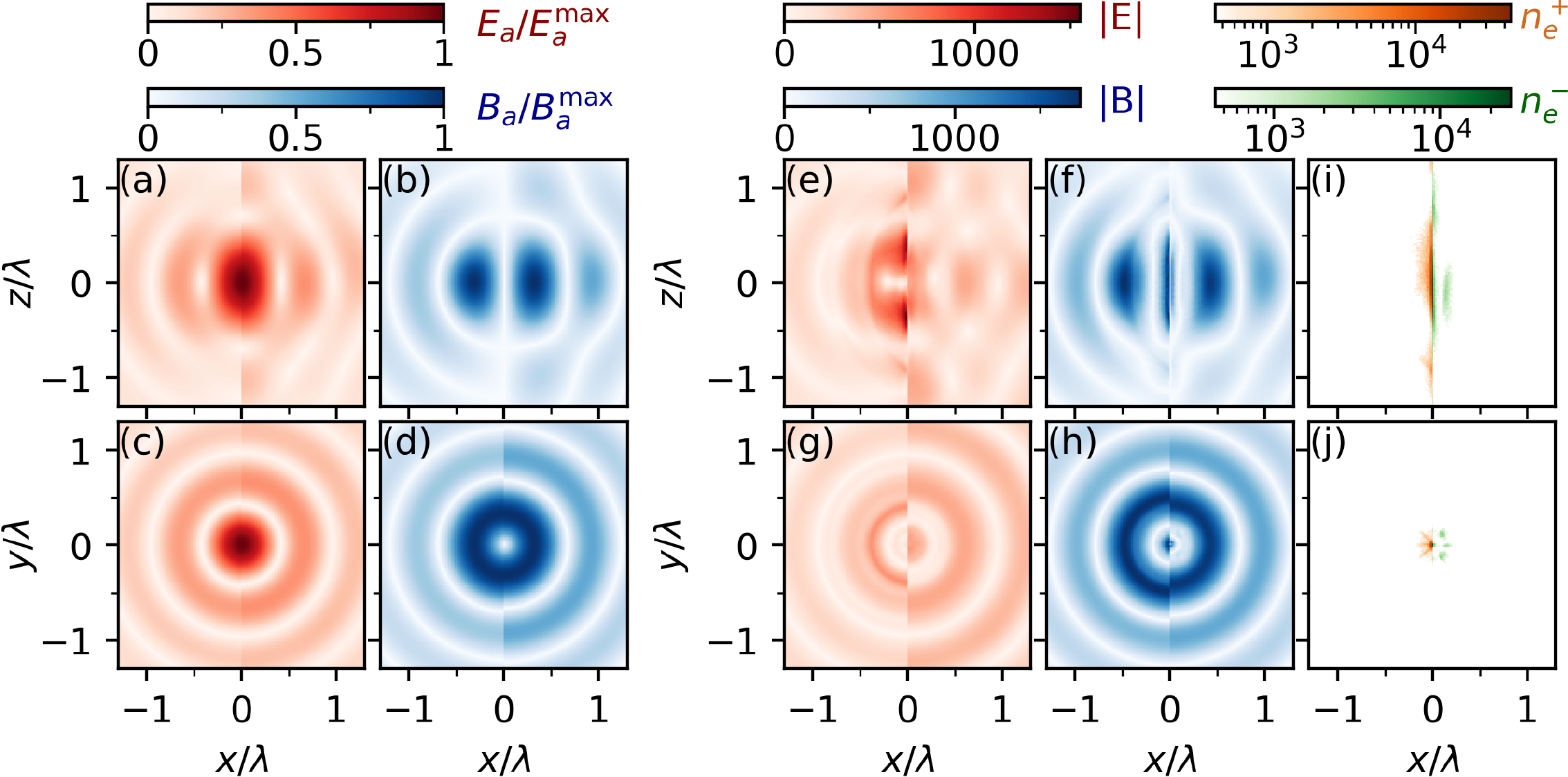}
	\caption{\label{fig:struct12} Field and plasma distribution in the case of target irradiation by 12 tightly focused laser beams configured as e-dipole wave. Electric (a), (c) and magnetic (b), (d) field distributions in vacuum (without any targets) normalized by their maxima are compared in $xz$ planes (a), (b) and $xy$ planes (c), (d) in case of ideal e-dipole wave (left half of the panels) and in case of 12 laser beams (right half of the panels). Comparison of the first basic plasma-field structures formed in cases of ideal e-dipole wave (left half of the panels) and of 12 beams (left half of the panels) irradiating a spherical target with $R_0=2\mu$m and initial density $n_0=1$ is presented in panels (e)-(j). Electric (e), (g) and magnetic (f), (h) field distributions as well as distributions of electrons and positrons (i), (j) are demonstrated in $xz$ planes (e), (f), (i) and in $xy$ panels (g), (h), (j). For illustrative purposes, keeping in mind equal quantity of electrons and positrons in the first basic structure, in (i) and (j) the left panel shows positron distribution while the right one demonstrates electron distribution.}
\end{figure*}

Although Eq.~(\ref{Eq:breakcondmin}) is transcendental, the logarithmic term allows estimating both roots. When the maximum threshold is reached, the quantity of electrons $f_{drag}N_{vb}$ gathered in focus should approach the critical quantity of electrons in focus $N_{cth}$ and the logarithmic term can be neglected in comparison with $\tau$ especially if $\Gamma>1/T$. So, the above estimated  maximum threshold initial electron quantity is 
\begin{equation}
    \label{Eq:Nvbmax}
    N_{vb}^\mathrm{max}\approx\Gamma_{esc}N_{cth}\tau\approx10^{12}.
\end{equation}

When the minimum threshold is reached, $N_{vb}^\mathrm{min}\ll N_{cth}$. Since the function $\log\left(\frac{1}{x}\right)$ rapidly increases as $x\rightarrow0$, the minimum root of Eq.~(\ref{Eq:breakcondmin}) is very close to the root of the equation $\tau-\log\left(\frac{N_{cth}}{f_{drag}N_{vb}}\right)/\Gamma=0$ which is 
\begin{equation}
    \label{Eq:Nvbmin}
    N_{vb}^\mathrm{min}\approx N_{cth}e^{-\Gamma\tau}/f_{drag}\approx1.9\times10^6.
\end{equation}
So, approximate analytical roots and numerical roots are almost the same. Moreover, the solutions Eq.~(\ref{Eq:Nvbmax}) and (\ref{Eq:Nvbmin}) show that both roots depend on $N_{cth}$ and pulse duration. The maximum root is caused, among others, by the rate of particle escape from focus $\Gamma_{esc}$, while the minimum root is determined by the growth rate of the QED cascade $\Gamma$ and the fraction of initial electrons $f_{drag}$ pushed to focus and gathered within it.

\section{\label{sec:multi}Multibeam setup}

\begin{figure*}
	\includegraphics[width=1\linewidth]{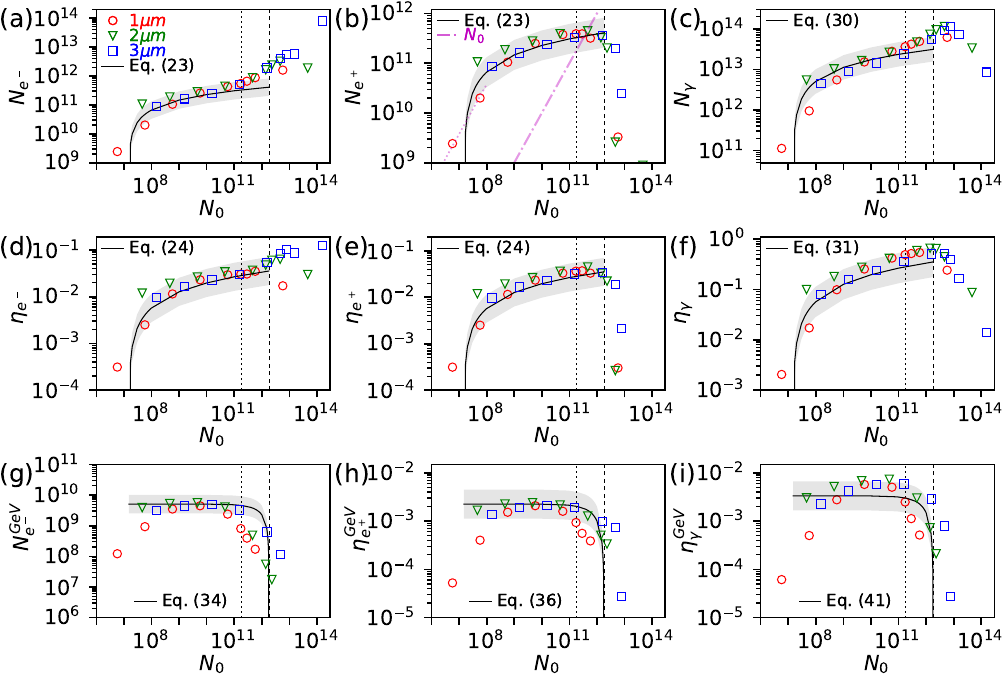}
	\caption{\label{fig:nsum12} Total quantity of electrons $N_{e^-}$ (a), positrons $N_{e^+}$ (b) and $\gamma$-photons $N_\gamma$ (c), and efficiencies of their generation $\eta_{e^-}$ (d), $\eta_{e^+}$ (e), $\eta_\gamma$ (f), respectively, due to irradiation of spherical targets with different initial electron quantity $N_0$ by tightly focused 12 beams configured as e-dipole. The total laser power is 30~PW and duration of each pulse is 15~fs. Quantity of GeV electrons $N_{e^-}^{GeV}$, efficiencies of generation of GeV positrons $\eta_{e^+}^{GeV}$ and GeV $\gamma$-photons $\eta_{\gamma}^{GeV}$ are presented in panels (g)-(i), respectively. Markers in all panels show results of simulations and their shapes (and colors) labeled in legends of panel (a) correspond to different target radii. The sloping dash-dotted line in (b) shows $N_{e^+}=N_0$ while the sloping dotted line connecting two points at $N_0<10^8$ corresponds to $N_{e^+}=350N_0$. The vertical dotted and dashed lines denote threshold initial electron quantities $\overline{N_{0th}}$ and $N_{0th}$ in accordance with Eqs.~(\ref{Eq:N0thl}) and (\ref{Eq:N0th}), respectively. The solid lines labeled by references to corresponding equations demonstrate approximations of numerical results. Grey color areas demonstrate the difference form estimates by no more than twice.}
\end{figure*}

In the previous section we considered the interaction of a laser wave with idealized field structure with targets. This section is devoted to the field configuration formed by a number of tightly focused laser beams, that is more relevant for possible experiments. The interaction of multibeam laser radiation with targets is a much more complex problem in view of plenty parameters. Here we will only partially touch on this problem in order to testify the robustness of the  interaction of a laser field configured as a dipole wave with a target. We consider two configurations consisting of 12 beams arranged in two belts and 6 beams arranged in one belt \cite{Gonoskov_PRL2014}, the foci of these beams are in the same position, the pulses are ideally synchronized and there are no  energy and power perturbations of these pulses. First, we will consider the case of 12-beam configuration that is more close to the ideal one and will then analyze how the results depend on the number of laser beams, their duration, power and polarization.

\subsection{12 laser beams}
For practice it is a key question whether the qualitative model of the irradiation of a target by an e-dipole wave developed above can be applied for the case of a multibeam setup. To check it let us first consider the configuration of 12 beams which is one of the best approximations of an ideal e-dipole wave. 

Since not only the total quantities of particles and $\gamma$-photons escaping the focus, but also the quantities of particles and photons with highest energies are of interest, we will first compare the field distributions in vacuum relevant for the linear stage of the QED cascade in cases of an ideal e-dipole wave and 12 beams focused in the form of an e-dipole wave. The focal field distributions in these cases are very similar (Fig.~\ref{fig:struct12}(a)-(d)). However, there are a few differences.

In the case of 12 beams, the amplitudes $E_a$ and $B_a$ are approximately 10\% less, but the spatial scale of the focal spot along the $z$ axis is approximately 10\% larger (Fig.~\ref{fig:struct12}(a), (b)). The scale of the electric field antinode region slightly decreases in the transverse direction (Fig.~\ref{fig:struct12}(c), (d)).  According to the simulations, with such field modifications $\gamma_{GeV}\approx1300$ is almost the same as in the idealized case (Eq.~(\ref{Eq:gamgev})), while the quantum parameter $\chi_{GeV}\approx3.48$ becomes 10\% less than the one in Eq.~(\ref{Eq:chigev}) due to the decrease in the field amplitudes. As a result, the rate $\Gamma\approx2.34/T$ of QED cascade development is 15\% less in comparison with $\Gamma$ (Eq.~(\ref{Eq:gammarate})) in the ideal e-dipole wave. Due to such a decrease at $N_0\lesssim10^8$, the quantity of particles in focus does not reach $N_{cth}$ and only the linear stage of the QED cascade is realized, this corresponds to  $N_{e^+}\propto N_0$, in Fig.~\ref{fig:nsum12}(b) the dotted sloping line is parallel to the dash-dotted line. Thus, an additional regime, {\it the linear regime of QED cascade} appears but it is not efficient for generating $\gamma$-photons or accelerated pairs because of a small quantity of pairs in focus. In this regime the fields are not affected by the generated pairs.

Also, the reduction of $\chi_{Gev}$ leads to a decrease in the rate of GeV photon emission $W_{rad}^{GeV}$. Using Eq.~(\ref{Eq:wradgev}) with relevant $\gamma_{GeV}$ and $\chi_{GeV}$ this rate is around $W_{rad}^{GeV}\approx0.11/T$. In turn, $\zeta_{av}^{GeV}\approx0.85$ according to Eq.~(\ref{Eq:zetagev}) is almost the same.

Another property of the multibeam configuration is that the fields decrease slowly with increasing distance from the focus, and the regions of field antinodes, especially in the vicinity of the $z$ axis around the focus, become more pronounced (Fig.~\ref{fig:struct12}(a), (b)). A slower field decay, more pronounced antinode regions around the focus, as well as an azimuthal-nonuniform phase front in the case of the multibeam setup result in a decrease in the quantity of electrons reaching the focus due to the initial laser compression of the target. In the case of 12 beams, $f_{drag}\approx0.05$ is approximately twice as small as in the idealized case. But it should be pointed out that azimuthal modulations of the fields in focus are weak (Fig.~\ref{fig:struct12}(c) and (d)).

Modifications of basic plasma-field structures are the consequence of the changes in the vacuum field structure. These modification are more quantitative than qualitative. We consider the differences using an example of the first basic plasma field structure (Fig.~\ref{fig:struct12}(e)-(j)).  Opposite to the case of the ideal dipole wave, in the case of a multibeam setup even weak inevitable azimuthal field modulations provoke much stronger seeds for the formation of current sheets. Consequently, the current sheets become much more prominent and a much smaller quantity of particles takes part in pinching (Fig.~\ref{fig:struct12}(i), (j)). This leads to a decrease in inductive magnetic and electric fields at the edges of the $e^-$-$e^+$ plasma column (Fig.~\ref{fig:struct12}(e), (f), (g), (h)). However, the amplitudes of the total fields (inductive plus wave) acting on plasma are almost equal to those in the idealized case. Current sheets are better separated from the plasma column and are at a larger distance from it (in other words from the $z$ axis).  According to the simulations, such field modifications result in a twice lower critical quantity of electrons in focus $N_{cth}\approx0.9\times10^{11}$. This quantity corresponds to the quasi-stationary quantity of pairs, which is also  maximal for the first an the second basic structures.  If the quantity of pairs becomes larger, the field penetration into plasma is suppressed, the laser compression of the target becomes weak, and the third basic plasma-field structure is formed.  

There is another consequence of the modifications of the basic structures. More prominent antinode regions around the focus are typical not only for the vacuum field distribution but for the field distribution of the first and the second basic structures as well (Fig.~\ref{fig:struct12}(e), (f)). Consequently, particles should have higher energy in order to overcome the barriers of these regions. Thus, the rate of escape decreases from 1.16/T (Eq.~(\ref{Eq:gammaesc})) in the idealized case to $\Gamma_{esc}\approx0.91/T$ in the case of 12 beams. In turn, these regions also reduce the fraction $f_{GeV}$ of particles having GeV energy from 0.16 to 0.14.

\begin{figure*}
	\includegraphics[width=1\linewidth]{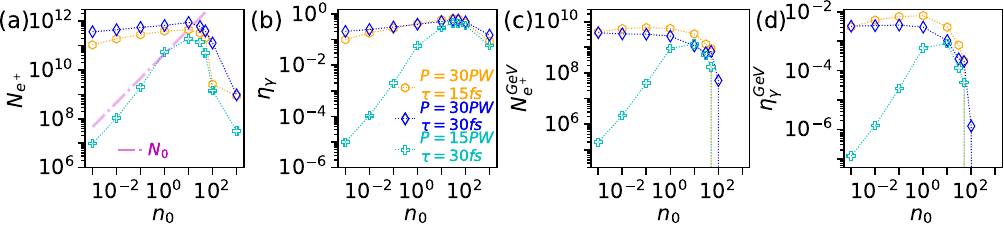}
	\caption{\label{fig:nsum_Pt} Comparison of the total quantity of escaping positrons (a), the total efficiency of $\gamma$-photon generation (b), the quantity of escaping GeV positrons (c) and the efficiency of GeV photon generation (d) as a function of initial target density for three configurations of laser radiation interacting with a spherical target having radius $2~\mu$m. The first one is 12-beam configuration with a total power of 30~PW and a pulse duration of 15~fs. The second configuration differs from the first one by longer pulse duration, 30~fs. The third configuration corresponds to 12 beams of total power 15~PW with duration 30~fs. Markers labeled in the panel (b) denote each field configuration. The dash-dotted line in (a) shows the initial quantity of electrons in the target.}
\end{figure*}

Nevertheless, we remind the readers that the basic structures are quite robust and the average particle energy weakly depends on power \cite{Efimenko_SR2018,Efimenko_PRE2019}. Our simulations demonstrate that this robustness remains even when the ideal wave is replaced by the 12-beam configuration. Since the field amplitudes are close in both cases, the difference between the particle energies in these cases and between the quantum parameters  vanishes. So, in the first and the second basic structures the values of $\gamma$ and $\chi$ can be estimated from Eqs.~(\ref{Eq:estgamma}) and (\ref{Eq:chival}), respectively. Consequently the rate of $\gamma$-photon emission $W_{rad}$ and the average energy fraction taken away by photons from particles $\zeta_{av}$ are given by Eqs.~(\ref{Eq:wrad}) and (\ref{Eq:zetaav}).

Since the differences between the 12-beam configuration and the ideal e-dipole wave is not so strong, the dependence of the quantities of the particles and $\gamma$-photons escaping the focus on $N_0$ is quite similar (Fig.~\ref{fig:nsum12}). Again, we can confirm that $N_0$ persists to be a similarity parameter up to $N_{0th}$ (Eq.~(\ref{Eq:N0th})) and the threshold values $N_{0th}$ and $\overline{N_{0th}}$ (Eq.~(\ref{Eq:N0thl})) differ in the  regimes of interaction and formation of different basic structures. Note that, since $N_{cth}$ and $f_{drag}$ are approximately twice as small as in the case of 12-beam configuration, the threshold initial densities $n_{0th}$, $\overline{n_{0th}}$ and the threshold initial quantities of target electrons $N_{0th}$ and $\overline{N_{0th}}$, distinguishing the formation of different basic field structures, remain the same as in Eqs.~(\ref{Eq:n0th}), (\ref{Eq:n0thl}), (\ref{Eq:N0th}), (\ref{Eq:N0thl}), respectively, corresponding to the e-dipole wave.

In {\it the regime of $e^-$-$e^+$ self-compression} GeV particles and photons are generated with the highest efficiency and have the best directivity. The maximum quantity of GeV $e^-$ and $e^+$ ($\approx6\times10^9$ or 1~nC, Fig.~\ref{fig:nsum12}(g)) and the efficiency of their generation ($\approx0.2\%$, Fig.~\ref{fig:nsum12}(h)) can be obtained for targets with $N_0\approx10^{10}$. The maximum quantity of $\gamma$-photons ($2\times10^{10}$) and the efficiency of their generation (0.7\%) can be reached at larger $N_0$ around $5\times10^{10}$ (Fig.~\ref{fig:nsum12}(i)). Opposite to the idealized case, in the 12-beam configuration the range of target parameters leading to the first basic structure becomes slightly narrower due to the linear regime of the QED cascade emerging at $N_0\lesssim10^8$

The second {\it regime of multicomponent plasma self-compression} (like in the case of the e-dipole wave) is more suitable for positron generation. Their maximum quantity (or maximum quantity of pairs) of about $4\times10^{11}$ (or about 50~nC) can be generated at $N_0\approx5\times10^{11}$ (about $\overline{N_{0th}}$) with 4\% efficiency (Fig.~\ref{fig:nsum12}(a), (b), (d), (e)). The estimates Eqs.~(\ref{Eq:Nposmax}) and (\ref{Eq:etaposmax}) demonstrate a reasonable accuracy: $N_{e^+}^\mathrm{max}\approx3.3\times10^{11}$ and $\eta_{e^+}^\mathrm{max}\approx3.3\times10^{11}$. Thus, with this $N_0$, almost each electron of the target produces a pair. Such equality means that the maximum threshold initial quantity of electrons sufficient for vacuum breakdown $N_{vb}^\mathrm{max}$ corresponds to this regime. Since $\Gamma_{esc}$ and $N_{cth}$ become lower, according to Eq.~(\ref{Eq:breakcondmin}) or (\ref{Eq:Nvbmax}), $N_{vb}^\mathrm{max}$ decreases proportionally and becomes $3.5\times10^{11}$. This value coincides with the numerical results for $R_0=1$ and $3~\mu$m, but for $R_0=2~\mu$m the numerically determined threshold is three times higher (Fig.~\ref{fig:nsum12}(b)). In turn, the minimum threshold $N_{vb}^\mathrm{min}$ due to the $\exp{(-\Gamma\tau)}$ factor from Eq.~(\ref{Eq:Nvbmin}) increases up to $1.5\times10^7$. So, the range of conditions sufficient for the vacuum breakdown partially narrows, but remains quite wide. Note that here $N_{vb}^\mathrm{min}$ is an upper estimate, so more rigorous investigations are needed.

The transition to the third {\it regime of weak laser compression of the target} at $N_0\approx N_{0th}$ is favorable for the generation of a maximum quantity of $\gamma$-photons (around $10^{14}$) with efficiency around 60\% (Fig.~\ref{fig:nsum12}(c), (f)). But the directivity of such a source is not narrow, the angular width is tens of degrees or almost uniform. Finally, in the third regime for targets with $N_0\gg N_{0th}$ the largest quantity of electrons can be generated ($8\times10^{13}$ or around 1~$\mu$C) with efficiency 12\% (Fig.~\ref{fig:nsum12}(a), (d)). This optimum is observed for a target with $R_0=3~\mu$m and $n_0=1000$.

We note that the optimal quantities and efficiencies are not less than half of the corresponding values in the idealized case or even very close to them. Moreover, all the estimates for the quantities of particles and photons (Eqs.~(\ref{Eq:Ne-e+N0}), (\ref{Eq:NgammaN0})), estimates for efficiencies of their generation (Eqs.~(\ref{Eq:etae-e+N0}), (\ref{Eq:etagammaN0})) including estimates for GeV particles and photon (Eqs.~(\ref{Eq:Neesc}), (\ref{Eq:Nphgev}), (\ref{Eq:etaeesc}), (\ref{Eq:etaphgev})) give a good agreement with numerical results (Fig.~\ref{fig:nsum12}) if the parameters of particle dynamics and photon generation determined above corresponding to the case of 12-beam configuration are used. So, the interaction considered is rather robust.

\begin{figure*}
	\includegraphics[width=1\linewidth]{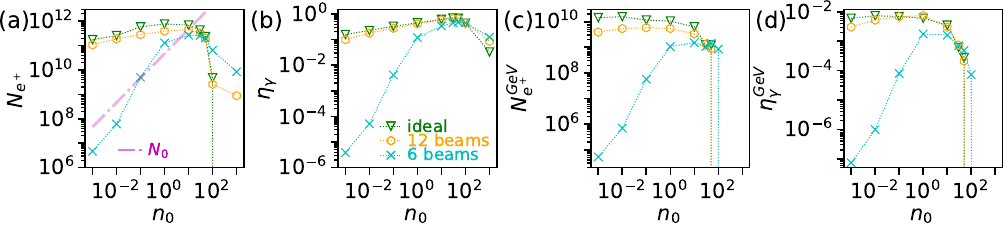}
	\caption{\label{fig:nsum_i_12_6} The same as in Fig.~\ref{fig:nsum_Pt} but for other laser field configurations: ideal e-dipole wave, 12 beams arranged in two belts and 6 beams arranged in one belt in order to mimick the e-dipole wave. Total laser power is 30~PW, pulse duration is 15~fs.}
\end{figure*}

However, in the multibeam configuration the accuracy of the zeroth approximation at $N_0<N_{0th}$, which assumes the dependence of laser-target irradiation primarily on $N_0$, is slightly worse. The ratio of the results of the interaction can be up to 2. The main reason for such variations is a finite pulse duration. Depending on the target radius, the field amplitude in focus may be stronger or weaker when the initial target is compressed. This results in larger or smaller quantities of the particles and $\gamma$-photons escaping the focus. This becomes crucially important when the quantity of pairs approaches the critical value $N_{cth}$, when the peak of the laser pulse is far from the focus or reaches it. This is especially important for GeV photons in the case of 12-beam configuration, because the reduction of $\chi_{GeV}$ brings the energy cut-off of photons closer to 1~GeV. 

Nevertheless, although this approximation is rough, it allows obtaining reasonable estimates of the quantities of particles and $\gamma$-photons in different energy ranges (as well as their generation efficiencies) and determining more accurately which parameters influence quantitative and energy characteristics of the interaction. We will use this instrument for analysis in the next section.

\subsection{The influence of power, duration, polarization and number of beams}
 
Although the considered 12-beam configuration is closer to the experimental realization, it is still more model than realistic. In experiment there are a lot of perturbations connected with coherent focusing of the number of beams, stability of pulse duration, polarization and power of each beam. In a word, there are plenty parameters of irradiation of a target by several tightly focused beams. Here, we will only touch on this problem and briefly discuss the importance of basic laser parameters: how the total power and pulse duration influence the interaction. Also, we consider the case of orthogonal beam polarization, when a combination of a number of beams reproduces an m-dipole wave, maximizing the magnetic field for the given power.

We consider as a reference the case of 12-beam configuration with a total power of 30~PW and a pulse duration of 15~fs (which reproduces well enough the case of the e-dipole wave) and compare the results in other field configurations with the ones presented above. In this section we will analyze  positron quantity and efficiency of $\gamma$-photon generation in different energy ranges, since these issues are the most intriguing in upcoming experiments at multipetawatt laser facilities. The results are presented for the spherical target with $R_0=2~\mu$m and with initial density $n_0$ from $10^{-3}$ to $10^3$.

Let us first discuss the case of a longer pulse duration of  30~fs, which is regarded to be the upper limit of the duration in the prospective XCELS facility. Since the field structure and field amplitudes remain the same as for a shorter pulse, only $f_{drag}$ in addition to $\tau$ is changed. As longer-pulse electrons pushed to focus are scattered earlier by a diverging wave (after passing the leading front through the focus) \cite{Bashinov_QE2013}, this partially prevents the electrons from reaching the focus, hence $f_{drag}$ decreases. 

According to Eqs.~(\ref{Eq:Ne-e+N0}) and (\ref{Eq:etagammaN0}), mainly the change in $\tau$ influences the results, because $f_{drag}$ is included in the argument of the logarithm. Thus, a longer duration allows increasing $N_{e^-,e^+}$ (Fig.~\ref{fig:nsum_Pt}(a) and (b)). Maximum $N_{e^-,e^+}$ becomes twice as large and reaches $\approx9\times10^{11}$ or $\approx100~n$C at $N_0\approx5\times10^{11}$ or $n_0=10$. Since the quantity of $\gamma$-photons is also twice as large and the laser energy is doubled, $\eta_{\gamma}\approx57\%$ (at $N_0=2\times10^{12}$) is close to $\eta_{\gamma}\approx65\%$ in the case of a twice shorter duration. However, the situation with GeV particles and photons is different. According to Eqs.~(\ref{Eq:Neesc}), $N_{e^-,e^+}^\gamma$ should slightly increase due to a decrease in $f_{drag}$, but this equation does not take the duration into account. In the case of a longer duration, for the same initial density within the considered range, the quantity of pairs in focus reaches the critical value when the peaks of the laser pulses are farther from the focus. So, at this moment of time the amplitude in focus is smaller, which negatively affects the high-energy part of the spectrum. Thus, in the considered range of $n_0$, the  quantity $N_{e^+}^{GeV}$ and efficiency $\eta_\gamma^{GeV}$ are smaller than the corresponding values in the case of a shorter pulse duration (Fig.~\ref{fig:nsum_Pt}(c) and (d)). In our simulations in the case of a longer duration, the maximum values of $N_{e^+}^{GeV}\approx3.7\times10^9$ and $\eta_{\gamma}^{GeV}\approx0.3\%$ are reached at the low boundary of the considered density range. We expect that the real maximum of these values depends less on the duration and should be at lower densities, which we have not considered.

\begin{figure*}
	\includegraphics[width=1\linewidth]{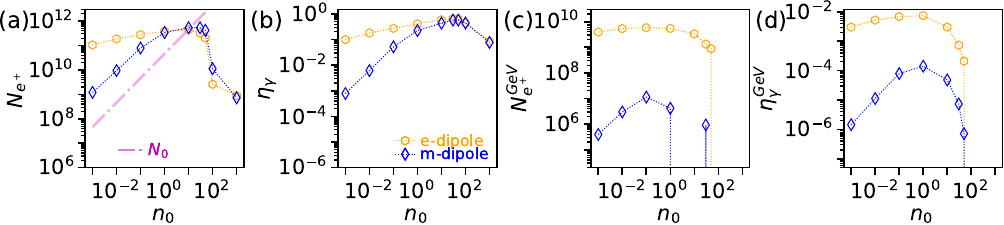}
	\caption{\label{fig:nsum_EB} The same as in Fig.~\ref{fig:nsum_Pt} but for other laser field configurations. The first one is 12 tightly focused beams configured as a e-dipole wave maximizing the electric field for the given power. The second configuration differs from the first one by orthogonal polarization, allows maximizing the magnetic field for the given power and mimicks the m-dipole wave. Total laser power is 30~PW, pulse duration is 15~fs.}
\end{figure*}

Another consequence of the longer pulse duration is a wider range of target parameters allowing vacuum breakdown. According to Eqs.~(\ref{Eq:Nvbmax}) and (\ref{Eq:Nvbmin}), $N_{vb}^{max}\approx8\times10^{11}$ (or maximum threshold density $n_{vb}^\mathrm{max}=17$) becomes almost twice larger. The minimum threshold initial electron quantity decreases much larger, due to the exponential factor in Eq.~(\ref{Eq:Nvbmin}).

Second, let us compare the case of a weaker power (15~PW) and a longer duration (30~fs) with the reference case keeping laser energy the same. Since the QED cascade is very sensitive to the field intensity, which becomes twice smaller, pair production is strongly suppressed because the average $\Gamma$ becomes less than $1/T$. As a result, at $N_0<5\times10^9$ ($n_0<0.1$) only {\it the regime of the linear QED cascade} is possible (Fig.~\ref{fig:nsum_Pt}). Thus, the range of $N_0$ corresponding to {\it the regime of self-compression of $e^-$-$e^+$ plasma} is narrowed. According to the estimates (\ref{Eq:Neesc}) and (\ref{Eq:etaphgev}), despite a smaller $\Gamma$, $N_{e^+}^{GeV}$ and $\eta_\gamma^{GeV}$ are almost an order of magnitude smaller (Fig.~\ref{fig:nsum_Pt}(c) and (d)) than the corresponding values in the reference case due to a noticeable decrease in the particle energy and consequently a decrease in $f_{GeV}$ and $W_{rad}^{GeV}$. The values of $f_{GeV}$ and $W_{rad}^{GeV}$ are very sensitive to the field amplitude, especially if the energy cut-off of particles or photons is close to 1~GeV.
{\it The regime of multicomponent plasma self-compression} is favorable for the maximum positron production and is initiated at the above laser parameters, but in the optimal case, maximum positron production is 3 times lower (Fig.~\ref{fig:nsum_Pt}(a)). The reduction of the maximum agrees with Eq.~(\ref{Eq:Nposmax}) due to $\Gamma<1/T$. The range of initial target densities sufficient for vacuum breakdown becomes very narrow for the same reason. According to our simulations (Fig.~\ref{fig:nsum_Pt}(a)), this range collapses to $0.6<n_0<1.6$ (or $3\times10^{10}<N_0<8\times10^{10}$).

However, since optimal total efficiency of $\gamma$-photon generation is reached at initial target densities (or initial quantities of electrons) close to the threshold of {\it the regime of weak target compression by laser radiation} and these photons are mainly generated by target electrons, this optimum is reduced much less and is equal to $\approx50\%$ (Fig.~\ref{fig:nsum_Pt}(b)). Thus, we can conclude that the reduction of power reduces the particle energy and the QED cascade rate and, therefore, significantly reduces the generation of GeV pairs, GeV $\gamma$-photons and leads to a decrease in the total quantity of pairs in this regime, respectively. 

Third, let us consider how a decrease in the number of laser beams influences the interaction. For this we compare the configuration with 6 laser beams arranged in one belt with a total power of 30~PW and a pulse duration of 15~fs with the reference 12-beam configuration. A smaller number of beams leads to the change of several parameters. As the phase front differs stronger from the spherical phase front, the laser radiation pushes the particles to focus worse and $f_{drag}$ decreases.  The spatial scale of the focal spot increases significantly along the electric field. Since the field inhomogeneity becomes smaller, the rate of particle escape $\Gamma_{esc}$ reduces. Also, the field amplitudes slightly decrease, causing reduction of the quantum parameter $\chi_{GeV}$ and consequently $f_{GeV}$, $W_{rad}^{GeV}$ and average $\Gamma$. However, $N_{cth}$ almost does not change. 

As a result, similarly to the previous case, the range of target parameters corresponding to {\it the regime of $e^-$-$e^+$ plasma self-compression} becomes narrower. Mainly due to the reduction of $\Gamma_{esc}$ to a lesser degree due to the reduction of $f_{GeV}$ and $W_{rad}^{GeV}$, the maximum quantity $N_{e^+}^{GeV}$ and the maximum efficiency $\eta_{\gamma}^{GeV}$ are several times smaller than the corresponding values in the reference case (Fig.~\ref{fig:nsum_i_12_6}(c) and (d)). For the same reason, in {\it the regime of multicomponent plasma self-compression} the maximum $N_{e^+}$ decreases but not so strong (Fig.~\ref{fig:nsum_i_12_6}(a)). Also, like in the previous case, the maximum total efficiency of $\gamma$-photon generation reduces only slightly, because $\gamma$-photons can be effectively generated by electrons from the target (Fig.~\ref{fig:nsum_i_12_6}(b)).

The configuration with 6 beams allows triggering vacuum breakdown but the range of densities sufficient for this is not so large: $0.1<n_0<6$ or $5\times10^9<N_0<3\times10^{11}$ (Fig.~\ref{fig:nsum_i_12_6}(a)). The minimum threshold of the vacuum breakdown increases because $f_{drag}$ and $\Gamma$ decrease. The maximum threshold reduces since the particles escape the focus slower.

It should be emphasized that a decrease in the number of beams is quantitatively similar to a decrease in the total power. In other words, in the case of a smaller number of laser beams, an increase in the total power allows reaching highest efficiencies and fluxes of pairs.

We have considered above the field configurations reproducing the e-dipole wave structure, maximizing the electric field for the given total power. Finally, let us analyze the results in the case of orthogonal polarization of laser beams and compare them with the results in the case of the reference configuration. The total power is 30~PW, pulse duration is 15~fs, the interference of 12 tightly focused laser beams in focus forms a toroidal structure of electric field and a poloidal structure of magnetic field. Thus, this configuration mimicks the structure of the m-dipole wave, maximizing the magnetic field for the given power. The particle and plasma dynamics in the m-dipole wave differs from that in the e-dipole wave and has unique advantages which can be useful in practice \cite{Bashinov_PRE2022,Efimenko_PRE2022}.

According to Ref.~\cite{Gonoskov_PRA2012}, the amplitude of electric field in the focus of the m-dipole configuration is approximately a half of the corresponding amplitude in the e-dipole configuration. This leads to several consequences. The QED cascade growth is slower for the m-dipole wave, for this reason the linear regime of the QED cascade is possible up to the initial densities $n_0=0.1$. Therefore, at $n_0<0.1$ the quantity of accelerated positrons escaping the focus and the total efficiency of $\gamma$-photon generation are much less than the ones in the case of the e-dipole configuration. However, at higher initial densities the results are almost the same. Since positrons and $\gamma$-photons are generated in processes which are influenced not only by energy but also by the quantum parameters $\chi$ (see Eq.~(\ref{Eq:chi})), both electric and magnetic field determine their probability. The electric field accelerates particles, while the transverse (relative to the particle momentum) field, primarily the magnetic field, determines $\chi$. Since the product of maximum electric and magnetic field amplitudes is equal in both field configurations, the optimal outcomes of positrons and  $\gamma$-photons should be comparable. This is confirmed by the results presented in Fig.~\ref{fig:nsum_EB}(a) and (b) at $n_0>1$.

However, electric field reduction shifts the spectra of particles and photons to lower energies, which significantly (by two orders of magnitude or even stronger) decreases the quantity of GeV positrons and the efficiency of generation of $\gamma$-photons (Fig.~\ref{fig:nsum_EB}(c) and (d)). 

Thus, we can conclude that the m-dipole configuration can be as efficient as the e-dipole configuration in the sense of the total quantity of generated pairs and $\gamma$-photons, but within a narrower range of target parameters. At the same time, the m-dipole configuration is not optimal in term of the generation of GeV pairs and photons. 

\section{\label{sec:con}Conclusion}

We have considered the interaction of laser radiation configured as a dipole wave with a power of several tens of PW and femtosecond duration with targets of different sizes and densities. Based on the analysis of plasma response we have given a qualitative description of this interaction. We distinguished four regimes of the interaction ({\it the linear regime of QED cascade}, {\it $e^-$-$e^+$ plasma self-compression}, {\it multicomponent plasma self-compression}, {\it weak laser compression of target}) and  three quasi-stationary basic plasma-field structures which these regimes can result in. We have proved that, for targets with sizes of about the wavelength, the main parameter or the similarity parameter determining the regime of interaction is the total quantity of electrons in the target $N_0$. The similarity parameter $N_0$ can significantly simplify selection of an appropriate target by varying its material and size in order to approach the required value of $N_0$. The threshold values of $N_0$ for different regimes have been identified. Note that the obtained results can be generalized to targets of other shapes and sizes if the quantity of electrons gathered by laser pulses in focus is considered as the similarity parameter. We have demonstrated this for nanowire targets.

Moreover, we showed that each regime is characterized by its own quantities of accelerated electrons, positrons and $\gamma$-photons which can be used in practice. The maximum quantity of accelerated positrons or pairs escaping from the focus is about $10^{12}$ (around 100~nC), they are generated with efficiency of about 8\% at $N_0$ corresponding to the transition from {\it the regime of $e^-$-$e^+$ plasma self-compression} to {\it the regime of multicomponent plasma self-compression}. The maximum quantity of $\gamma$-photons is about $10^{14}$, they are generated with efficiency of about 70\% and propagate almost quasi-uniformly at $N_0$ corresponding to the transition to {\it the regime of weak laser compression of the target}. This regime should be studied more rigorously because in this regime maximum quantity of accelerated electrons is produced. Their quantity can approach $10^{14}$  (around $10~\mu$C)  and the generation efficiency reaches 10\%, but average energy amounts to about several tens of MeV and these electrons propagate quasi-uniformly. In turn, the maximum quantities of particles and photons with maximum energy (above 1~GeV) propagating with optimal directivity (angular width around 10 degree) are reached in {\it the regime of $e^-$-$e^+$ plasma self-compression}. In this regime it is possible to generate and accelerate up to $10^{10}$ GeV pairs and photons with efficiencies up to 0.7\% in a wide range of target parameters.

We confirmed that such interaction and the resulting outcomes of particles and photons are quite robust. We demonstrated that in the case of the ideal e-dipole wave and 12-beam configuration mimicking such an ideal wave the results are very close. Based on the obtained estimates of the quantities of accelerated particles and photons, of the efficiencies of their generation we also showed that a decrease in power and in the number of beams lead to a strong decrease in the production of GeV particles and photons, while maximum total quantities of $\gamma$-photons reduce slightly. The production of positrons or pairs in this case also decreases. In turn, an increase in pulse duration almost proportionally increases the total production of accelerated particles and $\gamma$-photons, while the production of GeV particles and photons decreases due to the finiteness of laser pulses and their smooth temporal envelope.

Since the vacuum breakdown plays a significant role in such interactions, we determined the target parameters sufficient to trigger it. We showed that there are minimum and maximum threshold initial densities or initial quantities of target electrons. Both the minimum and maximum thresholds are determined by the field structure in focus and pulse duration. In addition, the minimum threshold strongly depends on the QED cascade growth rate and the possibility of laser pulses to gather electrons in focus, while the maximum threshold depends on the rate of particle escape from the focus.

Finally, we emphasize that a laser setup determines the maximum quantity of accelerated $e^-$, $e^+$ and $\gamma$-photons as well as a fundamental possibility of vacuum breakdown, while targets determine whether this maximum will be reached and whether vacuum breakdown will be triggered. We have shown that extreme laser-matter interaction can be controlled by choosing one target similarity parameter, which can significantly simplify optimization of upcoming breakthrough experiments at multipetawatt laser facilities.

\begin{acknowledgments}
The authors acknowledge the support of the Ministry of Science and Higher Education of the Russian Federation, Agreement No. 075-15-2021-1361 (A.V., A.A., A.V.), project FSWR-2023-0034 (V.V., I.M.). The authors acknowledge the use of the computational resources provided by the Joint Supercomputer Center of the Russian Academy of Sciences and the Lobachevsky University.
\end{acknowledgments}

\section*{Data Availability Statement}
Data available on request from the authors
\nocite{*}
\bibliography{main}

\end{document}